\documentclass[%
superscriptaddress,
longbibliography,
twocolumn,
amsmath,
amssymb,
aps,
prl,
floatfix,
10pt,
]{revtex4-2}

\usepackage{graphicx}
\usepackage{dcolumn}
\usepackage{bm}
\usepackage{siunitx}
\usepackage{xcolor}
\usepackage[colorlinks=true, linkcolor=blue, urlcolor=blue, citecolor=blue]{hyperref}
\usepackage{url}

\begin{document}

\title{
First-Principles Framework for the Prediction of Intersystem Crossing Rates in Spin Defects: The Role of Electron Correlation
}

\author{Yu Jin}
\affiliation{Pritzker School of Molecular Engineering, University of Chicago, Chicago, Illinois 60637, USA}
\author{Jinsoo Park}
\affiliation{Pritzker School of Molecular Engineering, University of Chicago, Chicago, Illinois 60637, USA}
\author{Marquis M. McMillan}
\affiliation{Pritzker School of Molecular Engineering, University of Chicago, Chicago, Illinois 60637, USA}
\author{Daniel Donghyon Ohm}
\affiliation{Department of Physics and Astronomy, Seoul National University, Seoul 08826, Korea}
\author{Corrie Barnes}
\affiliation{Pritzker School of Molecular Engineering, University of Chicago, Chicago, Illinois 60637, USA}
\author{Benjamin Pingault}
\affiliation{Q-NEXT, Argonne National Laboratory, Lemont, Illinois 60439, USA}
\affiliation{Materials Science Division, Argonne National Laboratory, Lemont, Illinois 60439, USA}
\affiliation{Pritzker School of Molecular Engineering, University of Chicago, Chicago, Illinois 60637, USA}
\author{Christopher Egerstrom}
\affiliation{Pritzker School of Molecular Engineering, University of Chicago, Chicago, Illinois 60637, USA}
\affiliation{Materials Science Division, Argonne National Laboratory, Lemont, Illinois 60439, USA}
\author{Benchen Huang}
\affiliation{Department of Chemistry, University of Chicago, Chicago, Illinois 60637, USA}
\author{Marco Govoni}
\affiliation{Department of Physics, Computer Science, and Mathematics, University of Modena and Reggio Emilia, Modena, 41125, Italy}
\affiliation{Materials Science Division, Argonne National Laboratory, Lemont, Illinois 60439, USA}
\affiliation{Pritzker School of Molecular Engineering, University of Chicago, Chicago, Illinois 60637, USA}
\author{F. Joseph Heremans}
\affiliation{Pritzker School of Molecular Engineering, University of Chicago, Chicago, Illinois 60637, USA}
\affiliation{Q-NEXT, Argonne National Laboratory, Lemont, Illinois 60439, USA}
\affiliation{Materials Science Division, Argonne National Laboratory, Lemont, Illinois 60439, USA}
\author{David D. Awschalom}
\email{awsch@uchicago.edu}
\affiliation{Pritzker School of Molecular Engineering, University of Chicago, Chicago, Illinois 60637, USA}
\affiliation{Q-NEXT, Argonne National Laboratory, Lemont, Illinois 60439, USA}
\affiliation{Materials Science Division, Argonne National Laboratory, Lemont, Illinois 60439, USA}
\affiliation{Department of Physics, University of Chicago, Chicago, Illinois 60637, USA}
\author{Giulia Galli}
\email{gagalli@uchicago.edu}
\affiliation{Pritzker School of Molecular Engineering, University of Chicago, Chicago, Illinois 60637, USA}
\affiliation{Department of Chemistry, University of Chicago, Chicago, Illinois 60637, USA}
\affiliation{Materials Science Division, Argonne National Laboratory, Lemont, Illinois 60439, USA}

\date{\today}
\begin{abstract}
Optically active spin defects in solids are promising platforms for quantum technologies. Here, we present a first-principles framework to investigate intersystem crossing processes, which represent crucial steps in the optical spin-polarization cycle used to address spin defects. Considering the nitrogen-vacancy center in diamond as a case study, we demonstrate that our framework effectively captures electron correlation effects in the calculation of many-body electronic states and their spin-orbit coupling and electron-phonon interactions, while systematically addressing finite-size effects. We validate our predictions by carrying out measurements of fluorescence lifetimes, finding excellent agreement between theory and experiments. The framework presented here provides a versatile and robust tool for exploring the optical cycle of varied spin defects entirely from first principles.
\end{abstract}

\maketitle

{\textit{Introduction}}---Optically active spin defects in solids offer a promising platform for advancing quantum technologies~\cite{wolfowicz2021quantum}. Prototypical systems, such as the negatively charged nitrogen-vacancy (NV$^-$) center in diamond~\cite{walker1979optical,doherty2013nitrogen} and neutral divacancy centers in silicon carbide~\cite{son2020developing}, have shown potential applications in quantum sensing~\cite{schirhagl2014nitrogen,barry2020sensitivity}, communication~\cite{childress2013diamond,christle2017isolated,wolfowicz2017optical,anderson2022five}, and computation~\cite{weber2010quantum,waldherr2014quantum} by functioning as spin qubits. First-principles calculations have been instrumental in studying these spin defects~\cite{freysoldt2014first,alkauskas2016tutorial,dreyer2018first,gali2023recent}, in particular their optical spin-polarization cycle, which is critical for the initialization and readout of qubit states~\cite{thiering2017ab,thiering2018theory,bian2024theory}. While radiative transitions have been extensively studied using first-principles approaches~\cite{alkauskas2014first,jin2021photoluminescence,jin2022vibrationally}, nonradiative intersystem crossing (ISC) transitions remain less explored. A central challenge lies in accurately accounting for electron correlation effects, which are important for the description of many-body electronic states and their associated spin-orbit coupling (SOC) and electron-phonon (e-ph) interactions~\cite{gali2024challenges}.

In this Letter, we present a theoretical and computational framework to predict ISC rates in spin defects, highlighting the key role of electron correlation in the accurate description of the defects' many-body states, their SOC, and e-ph interactions. We focus on the NV$^-$ center in diamond as an exemplar system and present results for ISC rates as a function of temperature. We validate our theoretical predictions by measuring fluorescence lifetimes, achieving excellent agreement between theory and experiment.

\begin{figure}
    \centering
    \includegraphics[width=8cm]{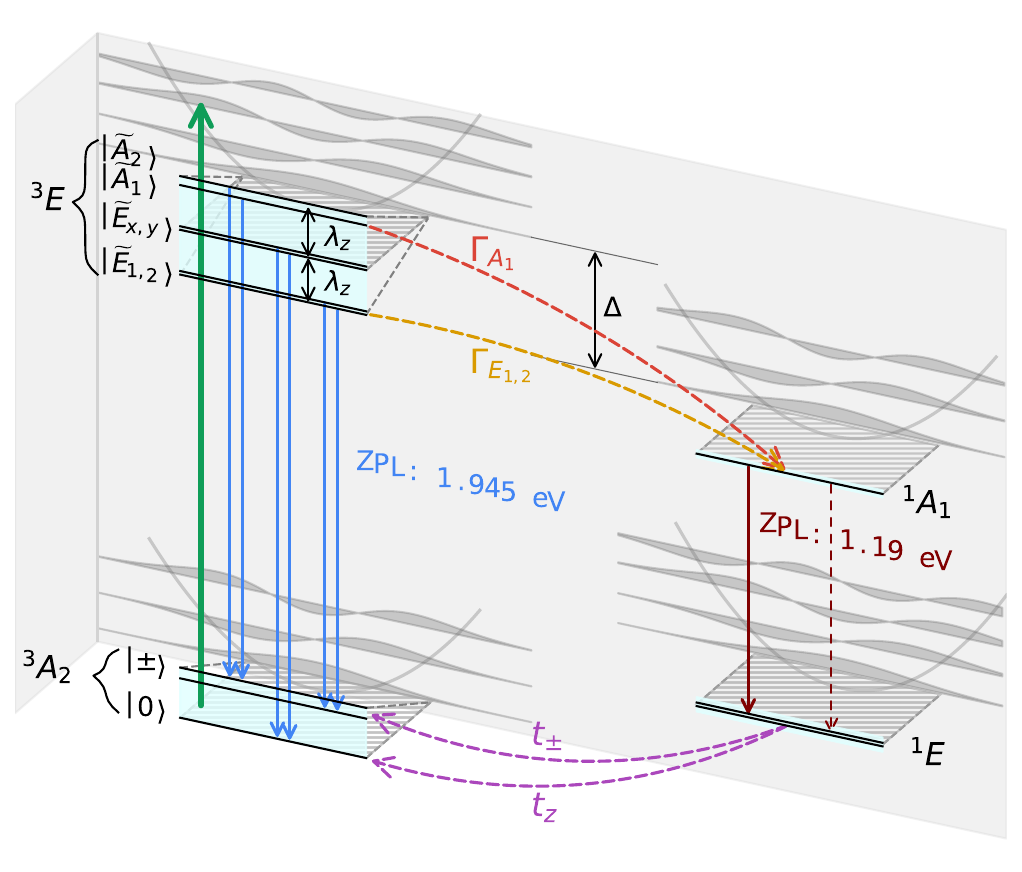}
    \caption{Many-body states and optical spin-polarization cycle of the NV$^-$ center in diamond. The spin-orbit splitting $\lambda_z$ represents the energy separation between the sublevels of the ${}^3\!E$ state. The intersystem crossing (ISC) rates from vibronic levels $\widetilde{A}_1$ and $\widetilde{E}_{1,2}$ to ${}^1\!A_1$ are labeled as $\Gamma_{A_1}$ (red dashed line) and $\Gamma_{E_{1,2}}$ (yellow dashed line). Radiative transitions ${}^3\!E \to {}^3\!A_2$ (blue solid line) and ${}^1\!A_1 \to {}^1\!E$ (maroon solid line), nonradiative transition ${}^1\!A_1 \to {}^1\!E$ (maroon dashed line), and ISC ${}^1\!E \to {}^3\!A_2$ (purple dashed line) are shown for completeness.}
    \label{fig:nv_description}
\end{figure}

The NV$^-$ center, comprising a nitrogen substitution and an adjacent vacancy, exhibits three localized defect orbitals ($a_1$, $e_x$, $e_y$) within the diamond band gap. In the negatively charged state, four electrons occupy these defect orbitals, 
resulting in a triplet ground state (${}^3\!A_2$), a triplet excited state (${}^3\!E$), and two singlet states (${}^1\!E$ and ${}^1\!A_1$)~\cite{manson2006nitrogen,maze2011properties,doherty2011negatively}. The optical spin-polarization cycle (Fig.~\ref{fig:nv_description}) consists of excitation from the triplet ground to the triplet excited state, followed by ISC to the higher singlet state, with relaxation to a lower singlet state, and finally an ISC process back to the triplet ground state. This cycle enables the initialization and readout of the qubit states, which are defined by the spin sublevels of the ${}^3\!A_2$ ground state~\cite{thiering2018theory}.

First-principles calculations have been successfully employed to investigate the optical transitions between ${}^3\!A_2$ and ${}^3\!E$~\cite{alkauskas2014first,razinkovas2021vibrational,jin2021photoluminescence} and between ${}^1\!A_1$ and ${}^1\!E$~\cite{thiering2018theory,jin2022vibrationally}. However, studies of ISC processes remain limited due to the challenges of accurately capturing electron correlation while also addressing finite-size effects. Previous ISC studies relying on mean-field approaches, e.g., density functional theory (DFT), overestimate measured ISC rates due to an inadequate description of SOC and e-ph interactions~\cite{thiering2017ab,thiering2018theory}. On the other hand, advanced quantum chemistry methods, applied to hydrogen-terminated carbon clusters simulating an NV$^-$ in diamond, underestimate ISC rates. The underestimate likely arises from quantum confinement effects and the negative electron affinity of small carbon clusters, which lead to inaccurate descriptions of electronic states and SOC~\cite{bhandari2021multiconfigurational,li2024excited,kundu2024designing}. Moreover, cluster models are unsuitable for the direct simulation of e-ph interactions in the bulk environment.

Here, we focus on ISC transitions from the $\widetilde{A}_1$ and $\widetilde{E}_{1,2}$ vibronic levels of ${}^3\!E$ to ${}^1\!A_1$ ($\Gamma_{A_1}$ and $\Gamma_{E_{1,2}}$ in Fig.~\ref{fig:nv_description}) and demonstrate the capabilities of our general theoretical and computational framework fully based on first-principles calculations. We combine several methods recently developed to investigate electronic structure properties beyond mean-field approaches. Specifically, we compute SOC using many-body wave functions from the quantum defect embedding theory (QDET)~\cite{ma2020quantum,ma2021quantum,vorwerk2022quantum,sheng2022green}, which accounts for electron correlation within the active space formed by defect states, while incorporating environmental effects through the screened Coulomb interaction. E-ph interactions are evaluated by computing the vibrational overlap function (VOF) between the ${}^3\!E$ and ${}^1\!A_1$ states, using atomic geometries and phonon modes obtained from spin-conserving and spin-flip time-dependent DFT (TDDFT) calculations~\cite{jin2023excited}. This method effectively accounts for the multiconfigurational nature of the ${}^3\!E$ and ${}^1\!A_1$ states and has been previously shown to accurately predict the absorption spectrum between singlet states of the NV$^-$ center~\cite{jin2022vibrationally}. Additionally, the dynamic Jahn-Teller (DJT)~\cite{bersuker2006jahn} and Herzberg-Teller (HT) effects~\cite{herzberg1933schwingungsstruktur,lin1974study} are incorporated into the VOF calculations. Importantly, our framework relies exclusively on solid-state calculations and is scalable to systems containing hundreds to thousands of atoms, thereby enabling a systematic treatment of finite-size effects for all computed quantities. We first describe the various steps of the computational protocol, and then we compare fluorescence lifetimes with measured values.

{\textit{Theoretical framework}}---At temperatures near 0 K, the ISC rate $\Gamma_{A_1}$ is calculated using Fermi's golden rule:
\begin{equation}
    \Gamma_{A_1} = \dfrac{4\pi}{\hbar} |\lambda_\perp|^2 F_{A_1} \left( \Delta \right),
\label{eq:gamma_a1_final}
\end{equation}
where $\lambda_\perp = 1 / \sqrt{2} \left\langle {}^3\!E(A_1) \right| \hat{H}_{\mathrm{SO}} \left| {}^1\!A_1 \right\rangle$ is the SOC matrix element between the $A_1$ electronic sublevel of ${}^3\!E$ and ${}^1\!A_1$~\cite{maze2011properties,doherty2011negatively}. Another matrix element, $\lambda_z = \left\langle {}^3\!E(A_1) \right| \hat{H}_{\mathrm{SO}} \left| {}^3\!E(A_1) \right\rangle$, determines the splitting of the sublevels in ${}^3\!E$. The term $F_{A_1}(\Delta)$ is the VOF for the transition from the $\widetilde{A}_1$ vibronic level of ${}^3\!E$ to the various vibronic levels of ${}^1\!A_1$. The energy $\Delta$ represents the gap between $\widetilde{A}_1$ and the lowest vibronic level of ${}^1\!A_1$. A detailed derivation of Eq.~\eqref{eq:gamma_a1_final} is provided in Sec.~S5 of Supplemental Material (SM)~\cite{SM}. The ISC rate $\Gamma_{E_{1,2}}$ can be similarly obtained by using the VOF for the $\widetilde{E}_{1,2}$ vibronic level [$F_{E_{1,2}}(\Delta)$].

To properly include electron correlation effects, $\lambda_{z/\perp}$ are computed using many-body wave functions obtained from QDET calculations~\cite{ma2020quantum,ma2021quantum,vorwerk2022quantum,sheng2022green} and using the many-body SOC operator, defined as~\cite{neese2007advanced,ai2022efficient}
\begin{equation}
    \hat{H}_{\mathrm{SO}} = \sum_{i=1}^{N_{e,\mathcal{A}}} \hat{\mathbf{z}}_i \cdot \hat{\mathbf{s}}_i=\sum_{p q} \sum_{\sigma \sigma^{\prime}}\langle p| \hat{\mathbf{z}}|q\rangle\langle\sigma| \hat{\mathbf{s}}\left|\sigma^{\prime}\right\rangle \hat{c}_{p \sigma}^{\dagger} \hat{c}_{q \sigma'},
    \label{eq:many-body-soc}
\end{equation}
where $N_{e,\mathcal{A}}$ is the number of electrons in the QDET active space, and $\hat{c}^{\dagger}_{p\sigma}$ ($\hat{c}_{q\sigma'}$) are the creation (annihilation) operators for orbital $p$ ($q$) in spin channel $\sigma$ ($\sigma'$). The one-body SOC operators entering Eq.~\eqref{eq:many-body-soc} are defined as
\begin{equation}
    \hat{h}_{\mathrm{SO}} = \hat{\mathbf{z}} \cdot \hat{\mathbf{s}} = \sum_{I=1}^{N_{\mathrm{atom}}} \left( \hat{V}_{\mathrm{NL}, I}^{\mathrm{FR}} - \hat{V}_{\mathrm{NL}, I}^{\mathrm{SR}} \right),
\end{equation}
where $ \hat{V}_{\mathrm{NL}, I}^{\mathrm{FR}}$ and $\hat{V}_{\mathrm{NL}, I}^{\mathrm{SR}}$ are the nonlocal (NL) components of the fully relativistic (FR) and the scalar relativistic (SR) pseudopotentials, respectively, for atom $I$ in the supercell~\cite{bachelet1982relativistic,verstraete2008density,kim2018effects}. Finite-size effects are addressed by converging the values of $\lambda_{z/\perp}$ with supercell size.

\begin{figure}
    \centering
    \includegraphics[width=8.6cm]{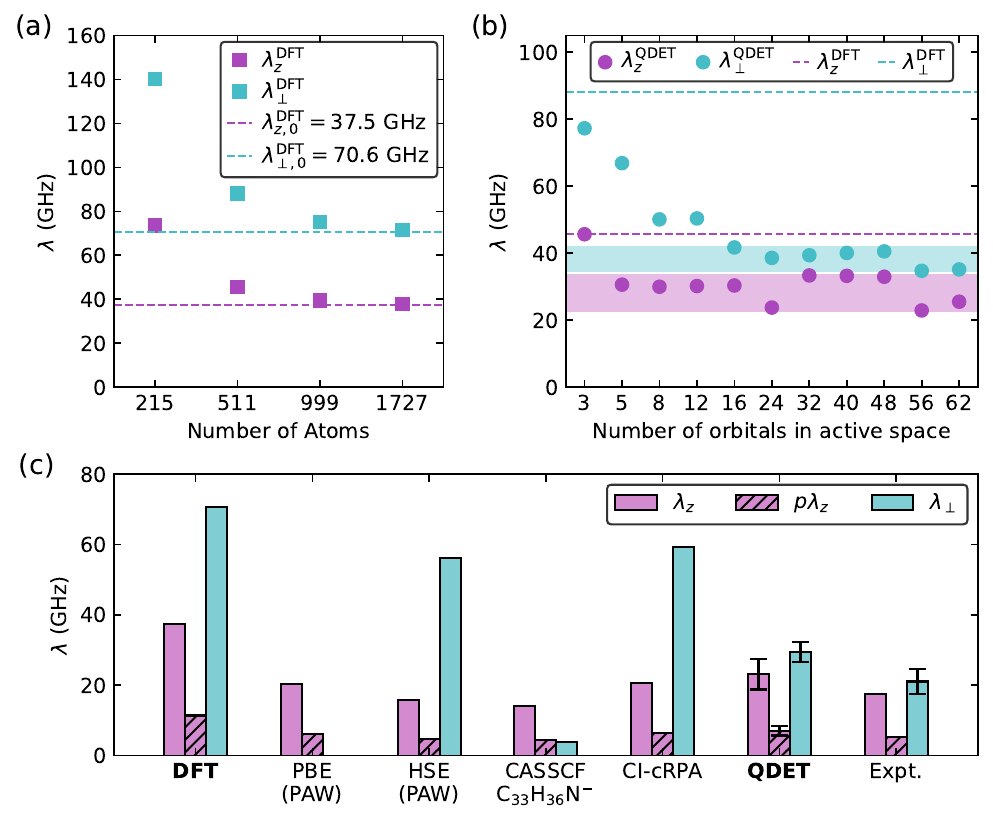}
    \caption{Computed spin-orbit coupling (SOC) parameters $\lambda_{z/\perp}$. (a) $\lambda_{z/\perp}^{\mathrm{DFT}}$ as a function of supercell size, with the extrapolated values ($\lambda_{z/\perp,0}^{\mathrm{DFT}}$) shown as dashed lines. $\lambda_{z/\perp}^{\mathrm{DFT}}$ is obtained using group-theory-derived many-body wavefunctions with a minimal defect orbital model ($a_1, e_x, e_y$). (b) $\lambda_{z/\perp}^{\mathrm{QDET}}$, computed using the many-body wavefunctions from QDET calculations in a 511-atom supercell with varying active spaces. The shaded areas indicate uncertainties due to active space selection. (c) Comparison of SOC parameters from this work (DFT and QDET) with previous theoretical studies~\cite{thiering2017ab,li2024excited,neubauer2024spin} and estimated experimental values~\cite{batalov2009low,bassett2014ultrafast,goldman2015state,goldman2017erratum,maze2011properties}. Full data available in Table~S3~\cite{SM}.}
    \label{fig:soc}
\end{figure}

The VOF $F_{A_1/E_{1,2}}(\Delta)$ is computed in two steps. First, contributions from $a_1$-type phonon modes, $F^{a_1}(\Delta)$, are calculated using the Huang-Rhys (HR) theory~\cite{huang1950theory,alkauskas2014first,jin2021photoluminescence}. This calculation uses the atomic geometries of ${}^3\!E$ and ${}^1\!A_1$ and the phonon modes of ${}^1\!A_1$, derived from spin-conserving and spin-flip TDDFT calculations~\cite{jin2022vibrationally}, which effectively include the multiconfigurational nature of these states. Next, the DJT effect of ${}^3\!E$, involving $e$-type phonon modes, is incorporated to obtain the final $F_{A_1/E_{1,2}}(\Delta)$~\cite{thiering2017ab}. The computed VOFs are extrapolated to the dilute limit using the force constant embedding approach~\cite{alkauskas2014first,razinkovas2021vibrational}. Equation~\eqref{eq:gamma_a1_final} assumes the validity of the Franck-Condon (FC) principle and treats $\lambda_\perp$ as independent of atomic vibrations. Here, we remove this assumption by accounting for the HT effect~\cite{herzberg1933schwingungsstruktur,lin1974study} and compute the derivative of $\lambda_\perp$ with respect to configuration coordinates in the calculation of $F^{a_1}(\Delta)$.

The energy gap $\Delta$ is approximated as the adiabatic energy difference between ${}^3\!E$ and ${}^1\!A_1$ at their respective equilibrium atomic geometries, assuming approximate cancellation of their zero-point energies. The vertical energy gap between ${}^3\!E$ and ${}^1\!A_1$ is computed using QDET, while their FC shifts (reorganization energies) are obtained from TDDFT~\cite{jin2022vibrationally,jin2023excited}. Further details on our computational framework are in Secs.~S1 and S2 of SM~\cite{SM}. Although not explored here, this computational framework is readily applicable to ISC transitions from ${}^1\!E$ to ${}^3\!A_2$ ($t_{\pm}$ and $t_{z}$ in Fig.~\ref{fig:nv_description}) by substituting the SOC matrix element, the VOF, and the energy gap with those between ${}^1\!E$ and ${}^3\!A_2$.

{\textit{Results}}---We first compute the SOC parameter, $\lambda_{z/\perp}$, using many-body wave functions derived from group theory, using the minimal basis set of defect orbitals $a_1$, $e_x$, and $e_y$ (see Tables~S1 and S2 in SM~\cite{SM}). The resulting expressions are -- $\lambda_z = |\langle e_x | \hat{\mathbf{z}} | e_y \rangle|/2$ and $\lambda_\perp = |\langle a_1 | \hat{\mathbf{z}} | e_x \rangle|/2$. Results obtained from these expressions are denoted as $\lambda_{z/\perp}^{\mathrm{DFT}}$. Figure~\ref{fig:soc}(a) shows the strong dependence of $\lambda_{z/\perp}^{\mathrm{DFT}}$ on supercell size, attributed to the overlap of defect orbitals with periodic images in small supercells. Extrapolating to the dilute limit via exponential decay yields $\lambda_{z,0}^{\mathrm{DFT}} = 37.5$ GHz and $\lambda_{\perp,0}^{\mathrm{DFT}} = 70.6$ GHz.

To properly account for electron correlation effects beyond the group theory analysis, we compute $\lambda_{z/\perp}$ using many-body wave functions from QDET calculations, denoted as $\lambda_{z/\perp}^{\mathrm{QDET}}$. Figure~\ref{fig:soc}(b) presents $\lambda_{z/\perp}^{\mathrm{QDET}}$ for a 511-atom supercell, highlighting the significant influence of electron correlation. While QDET wave functions resemble those from the group theory analysis, they do contain additional configurations that introduce small yet important corrections, leading to a notable decrease in $\lambda_{z/\perp}^{\mathrm{QDET}}$ compared to $\lambda_{z/\perp}^{\mathrm{DFT}}$ (details in Secs.~S3 and S4~\cite{SM} of SM). By accounting for the weak dependence of $\lambda_{z/\perp}^{\mathrm{QDET}}$ on the active space choice and applying a finite-size correction (assumed to be the same as for $\lambda_{z/\perp}^{\mathrm{DFT}}$), we determine the final SOC parameters as $\lambda_{z,0}^{\mathrm{QDET}} = 23.1 \pm 4.3$ GHz and $\lambda_{\perp,0}^{\mathrm{QDET}} = 29.4 \pm 2.8$ GHz. These values are used in the ISC rate calculations presented below.

In Fig.~\ref{fig:soc}(c), we compare our computed $\lambda_{z/\perp}$ with previous theoretical and experimental results. The quantity $p\lambda_z$ represents the measured fine structure splitting of ${}^3\!E$, where $p$ accounts for the Ham reduction factor due to the DJT effect. Using $p = 0.304$ from Ref.~\cite{thiering2017ab}, we obtain $p\lambda_{z,0}^{\mathrm{QDET}} = 7.0 \pm 1.3$ GHz, in good agreement with the measured values, $5.33\pm0.03$ GHz~\cite{batalov2009low,bassett2014ultrafast}, whereas $p\lambda_{z,0}^{\mathrm{DFT}}$ overestimates experiment by a factor of 2. Our computed $\lambda_{\perp,0}^{\mathrm{QDET}}$ also aligns well with the experimental estimate based on $\lambda_\perp / \lambda_z = 1.2 \pm 0.2$ relation~\cite{goldman2015state,goldman2017erratum}, while $\lambda_{\perp,0}^{\mathrm{DFT}}$ is approximately 3 times as large. We note that previous complete active
space self-consistent field (CASSCF) calculations~\cite{li2024excited} significantly underestimated $\lambda_\perp$, likely due to an inaccurate description of the NV$^-$ center's many-body wave functions with a small C$_{33}$H$_{36}$N$^-$ cluster.

\begin{figure}
    \centering
    \includegraphics[width=8.6cm]{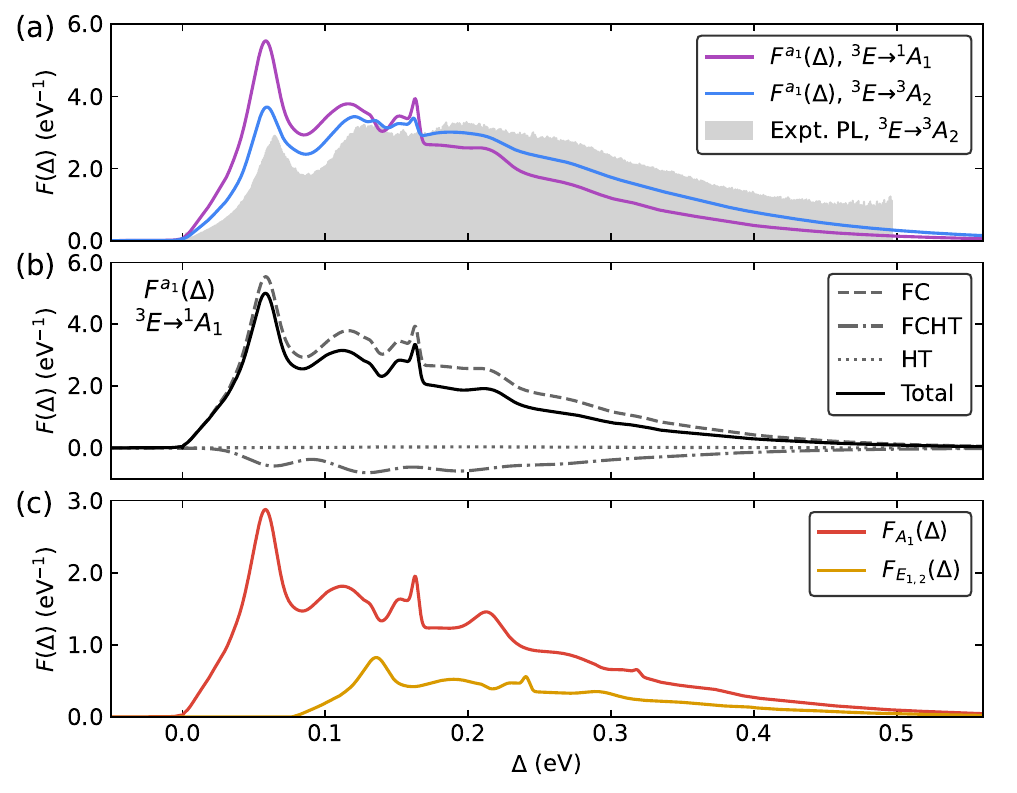}
    \caption{Vibrational overlap functions (VOFs). (a) Computed VOFs for the ${}^3\!E \to {}^1\!A_1$ and ${}^3\!E \to {}^3\!A_2$ transitions under the Franck-Condon (FC) principle, considering $a_1$-type phonon modes and extrapolated to the dilute limit. The experimental photoluminescence (PL) spectral function for ${}^3\!E \to {}^3\!A_2$~\cite{alkauskas2014first} is shown as the gray shaded area. (b) VOFs for the ${}^3\!E \to {}^1\!A_1$ transition with $a_1$-type phonon contributions, showing FC (dashed), Franck-Condon Herzberg-Teller (FCHT, dashed-dotted), and Herzberg-Teller (HT, dotted) components, with the total function as the solid line. (c) VOFs for $\Gamma_{A_1}$ and $\Gamma_{E_{1,2}}$ intersystem crossing transitions, incorporating the dynamic Jahn-Teller effect of the ${}^3\!E$ state.}
    \label{fig:spectral_fxn}
\end{figure}

As mentioned above, in our calculations of VOFs for the $\Gamma_{A_1}$ and $\Gamma_{E_{1,2}}$ ISC transitions, we first considered contributions from $a_1$-type phonon modes. The computed $F^{a_1}(\Delta)$ is shown in Fig.~\ref{fig:spectral_fxn}(a). Previous studies often derived $F^{a_1}(\Delta)$ from constrained-occupations DFT ($\Delta$SCF), which approximates ${}^3\!E$ and ${}^1\!A_1$ as single Slater-determinant states and inadequately account for electron correlation effects~\cite{thiering2017ab}; alternatively, $F^{a_1}(\Delta)$ was approximated by the VOF for the ${}^3\!E \to {}^3\!A_2$ photoluminescence (PL) transition~\cite{goldman2015state,goldman2017erratum,li2024excited}. Here, we compute $F^{a_1}(\Delta)$ using the HR theory~\cite{huang1950theory}, based on equilibrium atomic geometries of the ${}^3\!E$ and ${}^1\!A_1$ states and $a_1$-type phonon modes of ${}^1\!A_1$, obtained from spin-conserving and spin-flip TDDFT calculations. A scaling factor is applied to adjust the HR factors, correcting for underestimated atomic displacements due to the use of the Perdew-Burke-Emzerhof (PBE) functional~\cite{razinkovas2021vibrational,jin2022vibrationally} (see Sec.~S6 of SM~\cite{SM}). We find notable differences between $F^{a_1}(\Delta)$ for ${}^3\!E \to {}^1\!A_1$ and ${}^3\!E \to {}^3\!A_2$ transitions [Fig.~\ref{fig:spectral_fxn}(a)]. Specifically, in the range $\Delta \in [0.3, 0.4]$ eV, $F^{a_1}(\Delta)$ for ${}^3\!E \to {}^3\!A_2$ exceeds that of ${}^3\!E \to {}^1\!A_1$ by a factor of 1.5--1.9. Furthermore, the ${}^3\!E \to {}^3\!A_2$ PL line shape overestimates $F^{a_1}(\Delta)$ for ${}^3\!E \to {}^1\!A_1$ by a factor of 1.9--3.1 due to contributions from $e$-type phonon modes.

The VOF computed using the HR theory assumes the validity of the FC principle, where $\lambda_\perp$ is treated as constant between the atomic geometries of the initial and final states of the ISC transition. However, we find that $\lambda_\perp$ varies by 8\% between the ${}^1\!A_1$ and ${}^3E$ equilibrium atomic geometries (see Sec.~S7 of SM~\cite{SM}). This variation is accounted for by including the HT effect~\cite{herzberg1933schwingungsstruktur,lin1974study} in our calculations. Using the ${}^1\!A_1$ equilibrium atomic geometry as the reference, we compute the Franck-Condon Herzberg-Teller (FCHT) and HT VOFs, shown in Fig.~\ref{fig:spectral_fxn}(b). Our results reveal that while the HT term contributes only 1\%, the FCHT term reduces the FC term intensity by 19\%. These findings indicate that the HT effect has a minor influence on the VOF in the NV$^-$ center. However, our calculations demonstrate the capability of the framework developed here to assess vibronic coupling effects in SOC calculations~\cite{penfold2018spin}.

Finally, using an effective Hamiltonian describing the interaction between ${}^3\!E$ electronic states and $e$-type phonon modes~\cite{thiering2017ab}, we incorporate the DJT effect into the VOF calculations for $\Gamma_{A_1}$ and the $\Gamma_{E_{1,2}}$, yielding $F_{A_1}(\Delta)$ and $F_{E_{1,2}}(\Delta)$ [Fig.~\ref{fig:spectral_fxn}(c)]. The DJT effect mixes the $A_1$, $A_2$, and $E_{1,2}$ electronic levels of ${}^3\!E$, giving rise to the vibronic states $\widetilde{A}_1$, $\widetilde{A}_2$, and $\widetilde{E}_{1,2}$ through coupling with $e$-type phonons. This coupling reduces the intensity of $F_{A_1}(\Delta)$ relative to the bare $F^{a_1}(\Delta)$ and generates a nonzero $F_{E_{1,2}}(\Delta)$, thereby enabling the $\Gamma_{E_{1,2}}$ ISC transition. The onset of $F_{E_{1,2}}(\Delta)$ at 78 meV confirm its $e$-type phonon-assisted nature, consistent with the model of Ref.~\cite{goldman2015state}. Further details are provided in Sec.~S5 of SM~\cite{SM}.

\begin{figure}
    \centering
    \includegraphics[width=8.5cm]{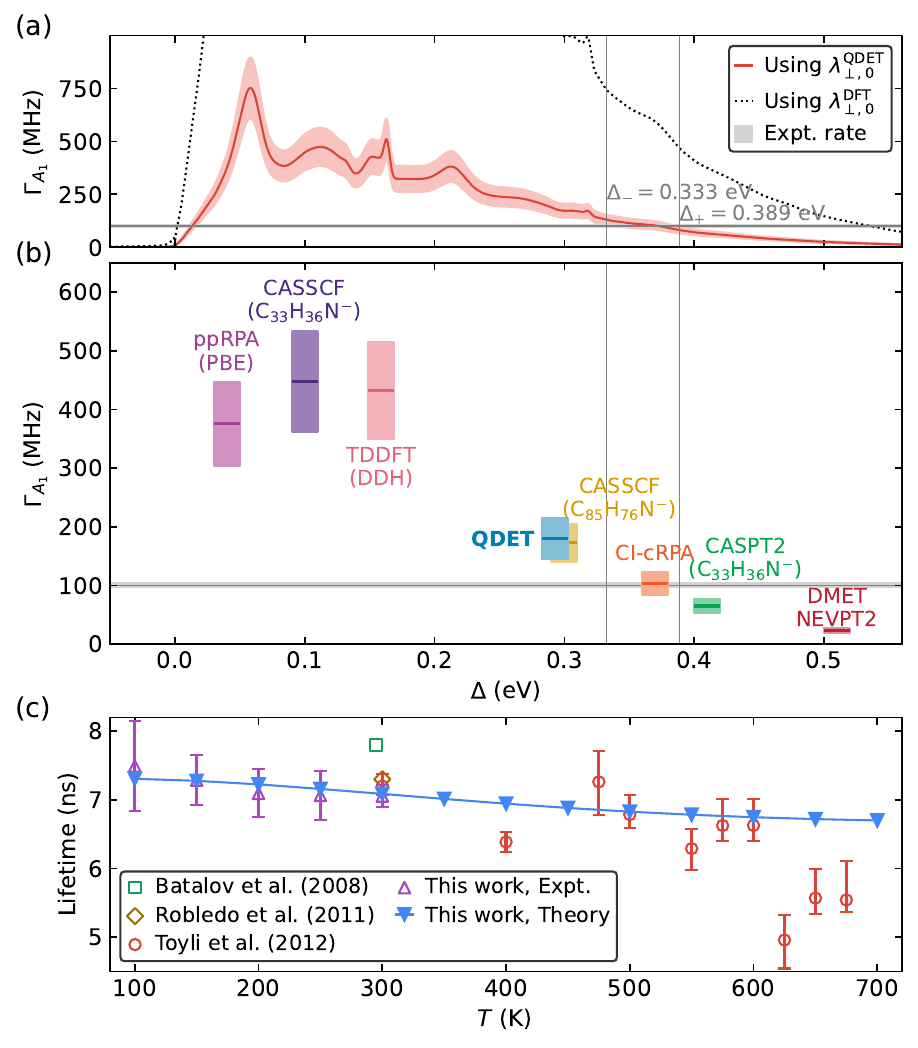}
    \caption{Intersystem crossing (ISC) rates and fluorescence lifetimes. (a) Computed ISC rate $\Gamma_{A_1}$ as a function of the energy gap ($\Delta$) between ${}^3\!E$ and ${}^1\!A_1$, with uncertainty (red shaded area) from the SOC parameter $\lambda_{\perp,0}^{\text{QDET}}$. The bounds $\Delta_-$ and $\Delta_+$ are determined by the intersection of the computed $\Gamma_{A_1}$ with experimental results~\cite{goldman2015phonon}. $\Gamma_{A_1}$ from $\lambda_{\perp,0}^{\text{DFT}}$ is shown for comparison (dotted line). (b) $\Gamma_{A_1}$ computed using $\lambda_{\perp,0}^{\text{QDET}}$ and the vibrational overlap function $F_{A_1}(\Delta)$ from this work, with $\Delta$ values from various theoretical methods~\cite{jin2023excited,chen2025qdet,haldar2023local,bockstedte2018ab,bhandari2021multiconfigurational,li2024excited,li2024accurate}, including Franck-Condon shifts from TDDFT~\cite{jin2023excited}. A visualization width of 0.02 eV is applied to each $\Delta$. Detailed data are provided in Table~S5 of SM~\cite{SM}. (c) Computed fluorescence lifetimes of the $|m_S| = 1$ sublevels of ${}^3\!E$ compared with experimental values from this work and previous studies by Batalov \textit{et al.}~\cite{batalov2008temporal}, Robledo \textit{et al.}~\cite{robledo2011spin}, and Toyli \textit{et al.}~\cite{toyli2012measurement}.}
    \label{fig:isc_rate}
\end{figure}

We now combine the calculations of the SOC parameter $\lambda_{\perp,0}^{\mathrm{QDET}}$ and the VOFs $F_{A_1} (\Delta)$ and $F_{E_{1,2}}(\Delta)$ to compute the ISC rates $\Gamma_{A_1}$ and $\Gamma_{E_{1,2}}$ as functions of the energy gap $\Delta$. Results for $\Gamma_{A_1}$ are shown in Fig.~\ref{fig:isc_rate}(a), while those for $\Gamma_{E_{1,2}}$ are in Fig.~S10 of SM~\cite{SM}. Comparing the computed $\Gamma_{A_1}$ with experimental data~\cite{goldman2015phonon,goldman2015state} yields an estimated $\Delta$ range of $[0.334, 0.389]$ eV. Similarly, the comparison of $\Gamma_{E_{1,2}}$ with experimental results~\cite{goldman2015phonon,goldman2015state} gives $[0.314, 0.375]$ eV. The overlap of these two ranges provides a refined estimate of $\Delta \in [0.334, 0.375]$ eV, offering a reliable prediction for the ${}^3\!E - {}^1\!A_1$ energy gap. The ISC rates computed using $\lambda_{\perp,0}^{\mathrm{DFT}}$, also shown in Fig.~\ref{fig:isc_rate}, overestimate experiments by roughly a factor of 6. The inverse dependence of $\Gamma_{A_1/E_{1,2}}$ on $\Delta$ suggests that the NV$^-$ center's ISC rates, and hence the optical spin-readout contrast, could be enhanced by decreasing $\Delta$, potentially through strain engineering~\cite{bhattacharyya2024imaging,wang2024imaging}.

Our QDET calculations predict a $\Delta$ value of 0.293 eV, which underestimates the lower bound of $\Delta$ shown in Fig.~\ref{fig:isc_rate}(a), resulting in a $\Gamma_{A_1}$ rate of $180\pm34$ MHz. This rate is higher than the measured value of $100.5\pm3.8$ MHz~\cite{goldman2015state}. We also compare previously reported $\Delta$ values from various theoretical approaches~\cite{jin2023excited,chen2025qdet,haldar2023local,bockstedte2018ab,bhandari2021multiconfigurational,li2024excited,li2024accurate} and compute the corresponding $\Gamma_{A_1}$ rates using $\lambda_{\perp,0}^{\textrm{QDET}}$ and $F_{A_1}(\Delta)$ from this work, as shown in Fig.~\ref{fig:isc_rate}(b). Notably, different theoretical approaches predict $\Delta$ in a broad range of $[0.04, 0.51]$ eV, leading to $\Gamma_{A_1}$ between 20 and 450 MHz. The values of $\Delta$ obtained from quantum chemistry approaches using carbon clusters are sensitive to both the level of theory and cluster size, resulting in $\Gamma_{A_1}$ variations by an order of magnitude. Overall, the comparison in Fig.~\ref{fig:isc_rate}(b) provides a valuable benchmark for quantitatively assessing the performance of various theoretical methods in predicting the energies of many-body states of spin defects. (See Sec.~S9 of SM~\cite{SM} for a detailed comparison of ISC rates evaluated here and in previous studies~\cite{goldman2015state,goldman2017erratum,thiering2017ab,li2024excited}).

Finally, we extend our computational framework for ISC rate calculations to finite temperatures ($T$) by summing contributions from different vibronic levels of ${}^3\!E$, weighted by the Boltzmann factor (see Sec.~S8 of SM~\cite{SM}). This approach effectively accounts for the phonon-induced orbital averaging effect and yields the averaged ISC rate from the $|m_S| = 1$ sublevels of ${}^3\!E$~\cite{goldman2015state}. We compute the fluorescence lifetimes of the $|m_S|=1$ sublevels of ${}^3\!E$ from 100 to 700 K, defined as $\tau (T) = 1 / \left[\Gamma_{\mathrm{Rad}} + \Gamma_{\mathrm{ISC}}(T) \right]$, where $\Gamma_{\mathrm{Rad}} = 82.9$ MHz is the radiative decay rate of ${}^3\!E$~\cite{goldman2015phonon}, and $\Gamma_{\mathrm{ISC}}(T)$ is the average ISC rate at a given $T$. To validate our calculations, we measured the fluorescence lifetimes of the $|m_S| = 1$ sublevels of ${}^3\!E$ from 100 to 300 K in 50 K intervals. Combining these with previous measurements from Batalov \textit{et al.}~\cite{batalov2008temporal}, Robledo \textit{et al.}~\cite{robledo2011spin}, and Toyli \textit{et al.}~\cite{toyli2012measurement}, we obtain experimental data spanning 100--600 K. Details on the experimental setup and lifetime fitting are provided in Secs.~S10 and S11 of SM~\cite{SM}. Despite moderate uncertainties in measurements and lifetime fitting, our computed lifetimes agree well with experimental values below 600 K, as shown in Fig.~\ref{fig:isc_rate}(c). At temperatures above 600 K, a significant decrease in lifetime is observed experimentally, which may be attributed to additional decay pathways~\cite{toyli2012measurement,goldman2015state} not yet included in our calculations, which will be explored in future work.

{\textit{Discussion}}---In summary, we have developed a general theoretical and computational framework for the evaluation of ISC rates in spin defects, enabling a complete description of their optical cycle based on high-level first-principles calculations. Using the NV$^-$ center in diamond as a case study, we have demonstrated that our framework accurately accounts for electron correlation in many-body states, thus enabling precise calculations of SOC parameters and e-ph interactions. We have also incorporated DJT and HT effects in VOF calculations, yielding ISC rates in good agreement with experiments. We have validated our approach at finite temperatures by comparing computed fluorescence lifetimes with experimental data, achieving excellent agreement between theory and experiment. Importantly, unlike cluster model-based methods~\cite{bhandari2021multiconfigurational,smart2021intersystem,lee2022spin,li2024excited}, our framework is entirely based on calculations for bulk systems and is scalable to systems containing hundreds to thousands of atoms, facilitating a systematic evaluation of finite-size effects. We emphasize that the framework presented here is general and applicable to studying ISC mechanisms and optical spin-polarization cycles in broad classes of spin defects; hence, it offers a robust tool for interpreting experiments under external fields, and guiding the engineering of spin defects, e.g., through the refinement of high-throughput discovery work flows.

Promising directions for future work include extending the current approach to include a multi-phonon-mode treatment of DJT and HT effects for the study of vibronic SOC~\cite{penfold2018spin,razinkovas2021vibrational,libbi2022phonon}. Additionally, our framework can be adapted to calculate phosphorescence rates for transition metal impurities in solids~\cite{diler2020coherent}, further expanding the applicability of first-principles calculations in predicting spin-related phenomena in quantum materials.


\textit{Acknowledgments}---The authors would like to thank Siyuan Chen, Victor Wen-zhe Yu, and Yuhang Ai for helpful discussions. The theoretical and computational work was primarily supported by the Midwest Integrated Center for Computational Materials (MICCoM) as part of the Computational Materials Sciences Program funded by the U.S. Department of Energy (Y.J., J.P., C.E., B.H., M.G., F.J.H., and G.G.). J.P. acknowledges the support from the Chicago Prize Postdoctoral Fellowship in Theoretical Quantum Science. Additional support for experimental validation includes the U.S. Department of Energy, Office of Science, Basic Energy Sciences Materials Sciences and Engineering Division (B.P. and D.D.A.), the SNU-Global Excellence Research Center establishment project, and the NRF grant funded by the Korean government (D.D.O.) (MSIT) No.~RS-2023-00258359, and from the AFOSR MURI under award No.~FA9550-23-1-0330 (M.M.M. and C.B.). This research used resources of the National Energy Research Scientific Computing Center (NERSC), a DOE Office of Science User Facility supported by the Office of Science of the U.S. Department of Energy under contract No.~DE-AC02-05CH11231 using NERSC Award No.~ALCC-ERCAP0025950, and resources of the University of Chicago Research Computing Center.

\textit{Data availability}---The data that support the findings of this Letter are openly available~\cite{data_statement}.


\bibliography{Main}

\end{document}



\title{Supplemental Material for ``First-Principles Framework for the Prediction of Intersystem Crossing Rates in Spin Defects: The Role of Electron Correlation"
}

\author{Yu Jin}
\affiliation{Pritzker School of Molecular Engineering, University of Chicago, Chicago, Illinois 60637, USA}
\author{Jinsoo Park}
\affiliation{Pritzker School of Molecular Engineering, University of Chicago, Chicago, Illinois 60637, USA}
\author{Marquis M. McMillan}
\affiliation{Pritzker School of Molecular Engineering, University of Chicago, Chicago, Illinois 60637, USA}
\author{Daniel Donghyon Ohm}
\affiliation{Department of Physics and Astronomy, Seoul National University, Seoul 08826, Korea}
\author{Corrie Barnes}
\affiliation{Pritzker School of Molecular Engineering, University of Chicago, Chicago, Illinois 60637, USA}
\author{Benjamin Pingault}
\affiliation{Q-NEXT, Argonne National Laboratory, Lemont, Illinois 60439, USA}
\affiliation{Materials Science Division, Argonne National Laboratory, Lemont, Illinois 60439, USA}
\affiliation{Pritzker School of Molecular Engineering, University of Chicago, Chicago, Illinois 60637, USA}
\author{Christopher Egerstrom}
\affiliation{Pritzker School of Molecular Engineering, University of Chicago, Chicago, Illinois 60637, USA}
\affiliation{Materials Science Division, Argonne National Laboratory, Lemont, Illinois 60439, USA}
\author{Benchen Huang}
\affiliation{Department of Chemistry, University of Chicago, Chicago, Illinois 60637, USA}
\author{Marco Govoni}
\affiliation{Department of Physics, Computer Science, and Mathematics, University of Modena and Reggio Emilia, Modena, 41125, Italy}
\affiliation{Materials Science Division, Argonne National Laboratory, Lemont, Illinois 60439, USA}
\affiliation{Pritzker School of Molecular Engineering, University of Chicago, Chicago, Illinois 60637, USA}
\author{F. Joseph Heremans}
\affiliation{Pritzker School of Molecular Engineering, University of Chicago, Chicago, Illinois 60637, USA}
\affiliation{Q-NEXT, Argonne National Laboratory, Lemont, Illinois 60439, USA}
\affiliation{Materials Science Division, Argonne National Laboratory, Lemont, Illinois 60439, USA}
\author{David D. Awschalom}
\email{awsch@uchicago.edu}
\affiliation{Pritzker School of Molecular Engineering, University of Chicago, Chicago, Illinois 60637, USA}
\affiliation{Q-NEXT, Argonne National Laboratory, Lemont, Illinois 60439, USA}
\affiliation{Materials Science Division, Argonne National Laboratory, Lemont, Illinois 60439, USA}
\affiliation{Department of Physics, University of Chicago, Chicago, Illinois 60637, USA}
\author{Giulia Galli}
\email{gagalli@uchicago.edu}
\affiliation{Pritzker School of Molecular Engineering, University of Chicago, Chicago, Illinois 60637, USA}
\affiliation{Department of Chemistry, University of Chicago, Chicago, Illinois 60637, USA}
\affiliation{Materials Science Division, Argonne National Laboratory, Lemont, Illinois 60439, USA}

\date{\today}

\maketitle

\tableofcontents
\newpage

\section{Computational Framework for ISC Rates Calculation}

\begin{figure*}
    \centering
    \includegraphics[width=12cm]{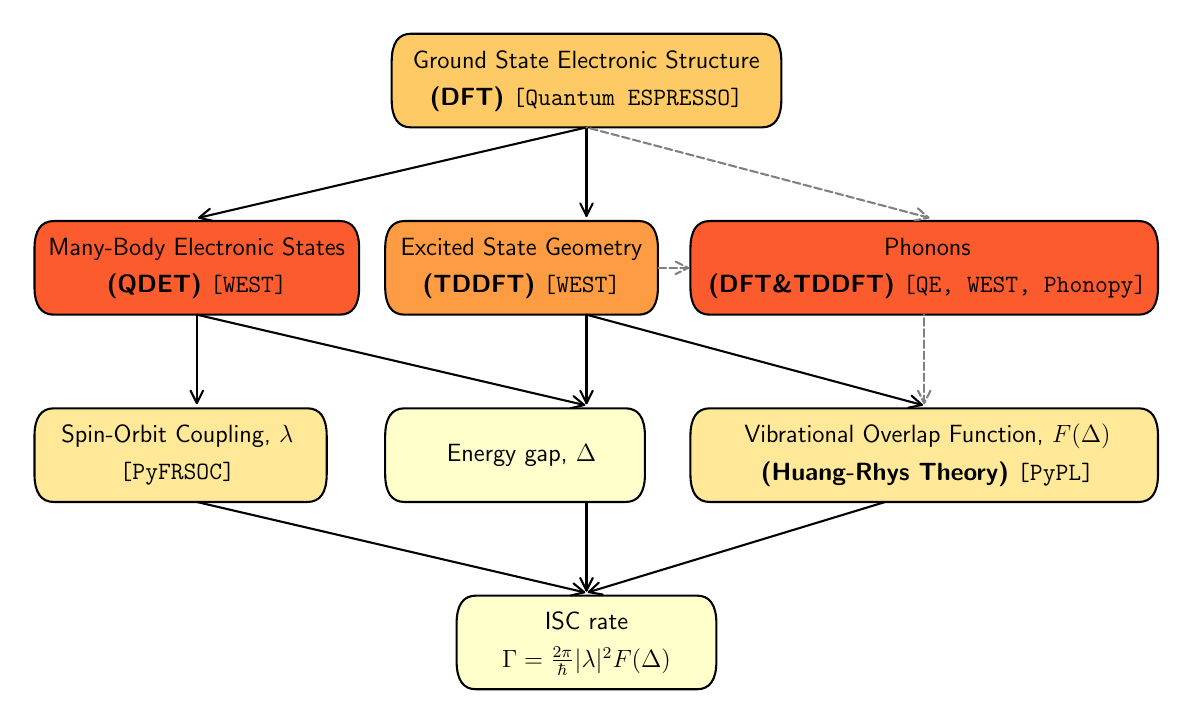}
    \caption{Computational workflow for the calculation of intersystem crossing (ISC) rates. Bold text within round brackets indicates the theories employed, with the corresponding codes specified in square brackets. Gray dashed arrows indicate procedures adopted in this work, which are, however, optional, as a one-dimensional approximation may be adopted (see text). The color gradient from light yellow to deep orange roughly indicates the increasing computational cost of each step.}
    \label{s-fig:workflow}
\end{figure*}

The computational workflow for calculating the intersystem crossing (ISC) rate is illustrated in Fig.~\ref{s-fig:workflow}. First, density functional theory (DFT) calculations are performed to obtain the ground state electronic structure using the \texttt{Quantum ESPRESSO (QE)} package~\cite{giannozzi2020qe,carnimeo2023quantum}. Next, quantum defect embedding theory (QDET) is employed using the \texttt{WEST} code~\cite{govoni2015large,yu2022gpu,ma2020quantum,ma2021qdet,sheng2022green,chen2025qdet} to compute many-body electronic states, which are then used to calculate spin-orbit coupling (SOC) matrix elements ($\lambda$) via the \texttt{PyFRSOC} package~\cite{jin2025pyfrsoc}.

At the same time, time-dependent DFT (TDDFT) calculations are conducted to optimize atomic geometries in the electronic excited states using the \texttt{WEST} code~\cite{jin2023excited}, and the results are used to compute the vibrational overlap function (VOF), $F(\Delta)$, based on the Huang-Rhys (HR) theory~\cite{huang1950theory} with the \texttt{PyPL} package~\cite{jin2021photoluminescence}. The vertical excitation energies from QDET, together with the Franck-Condon (FC) shifts obtained from TDDFT calculations, collectively yield the energy gap $\Delta$ between many-body states. The ISC rate is then determined using Fermi's golden rule, based on the computed values of $\lambda$, $\Delta$, and $F(\Delta)$.

For the calculation of $F(\Delta)$, an effective one-dimensional phonon approximation~\cite{li2024excited} may be applied, or a more comprehensive approach using all phonon modes, as used in this work, which requires additional frozen phonon calculations with the \texttt{Phonopy} package~\cite{phonopy-phono3py-JPCM,phonopy-phono3py-JPSJ}, as indicated by the gray dashed lines in Fig.~\ref{s-fig:workflow}. The additional calculations required for the treatment of dynamic Jahn-Teller (DJT) and Herzberg-Teller (HT) effects, carried out in this work, are omitted in Fig.~\ref{s-fig:workflow} for simplicity.

\section{First-Principles Electronic Structure Methods}

The electronic structure of the NV$^-$ center in diamond was calculated using DFT within the plane-wave pseudopotential framework, as implemented in the \texttt{Quantum ESPRESSO} package~\cite{giannozzi2020qe,carnimeo2023quantum}. The plane-wave energy cutoff was set to 60 Ry. We used the semilocal functional by Perdew, Burke, and Ernzerhof (PBE)~\cite{perdew1996generalized} and the optimized norm-conserving Vanderbilt (ONCV) pseudopotentials~\cite{hamann2013optimized,schlipf2015optimization,van2018pseudodojo}. The Brillouin zone of the supercell was sampled at the $\Gamma$ point. Supercells of various sizes were employed to investigate finite-size effects.

QDET calculations were performed using the \texttt{WEST} code~\cite{govoni2015large,yu2022gpu}. QDET uses $G_0W_0$ on top of a DFT starting point as the low-level electronic structure method to obtain the effective Hamiltonian parameters, incorporating an exact double-counting correction~\cite{sheng2022green}; the effective Hamiltonian is then solved with full configuration interaction (FCI) within a specified active orbital space~\cite{ma2020quantum,ma2021quantum}. The active spaces used in this work are denoted as $[(2N-2)e, No]$, where $(2N-2)$ electrons occupy $N$ orbitals. The $N$ orbitals were chosen as the the first $N$ orbitals below the valence band minimum of diamond. Once many-body wavefunctions were obtained, SOC matrix elements were computed as described in Eq.~(2) of the main text~\cite{neese2007advanced,ai2022efficient} using the \texttt{PyFRSOC} package~\cite{jin2025pyfrsoc}. The SOC operator was defined as the difference between the nonlocal terms of the full-relativistic and the scalar relativistic pseudopotentials.This approach goes beyond the commonly used lowest-order perturbation theory, as the fully relativistic pseudopotentials are constructed by directly solving the Dirac-type Kohn–Sham equation and retaining the large component of the four-component Dirac spinor. This approximation is accurate up to order $\frac{1}{c^2}$ due to normalization, where $c$ is the speed of light~\cite{theurich2001selfconsistent}, and achieves higher accuracy than the lowest-order perturbative methods, which is accurate up to order $\frac{Z^2}{c^2}$ with atomic number $Z$~\cite{verstraete2008density}, thus limiting their applicability to light elements. Additional details on the evaluation of the SOC matrix elements are provided in Sec.~\ref{s-sec:soc}.

The equilibrium atomic geometries of the ${}^3\!E$ and ${}^1\!A_1$ states were obtained by performing geometry optimization calculations, using the analytical nuclear forces in the spin-conserving and spin-flip TDDFT framework, within the Tamm-Dancoff approximation~\cite{hutter2003excited}, as implemented in the \texttt{WEST} code~\cite{jin2023excited}. Phonons modes of the ${}^1\!A_1$ state were computed using the frozen phonon approach, with configurations generated using the \texttt{Phonopy} package~\cite{phonopy-phono3py-JPCM,phonopy-phono3py-JPSJ} and a displacement of 0.01 \AA\ from the equilibrium atomic geometry. A 215-atom supercell was used for these calculations. To account for finite-size effects, the phonon modes were extrapolated to the dilute limit approximated by a $(12\times12\times12)$ supercell containing 13,823 atoms, following the force constant matrix embedding approach proposed by Alkauskas et al.~\cite{alkauskas2014first,razinkovas2021vibrational}. The VOFs were computed using the HR theory~\cite{huang1950theory}, based on the generating function approach implemented in the \texttt{PyPL} package~\cite{alkauskas2014first,jin2021photoluminescence}. Additional details are provided in Sec.~\ref{s-sec:spectral_fxn}.

\section{Electronic Structure of the Nitrogen-Vacancy Center in Diamond}

\begin{figure*}
    \centering
    \includegraphics[width=15cm]{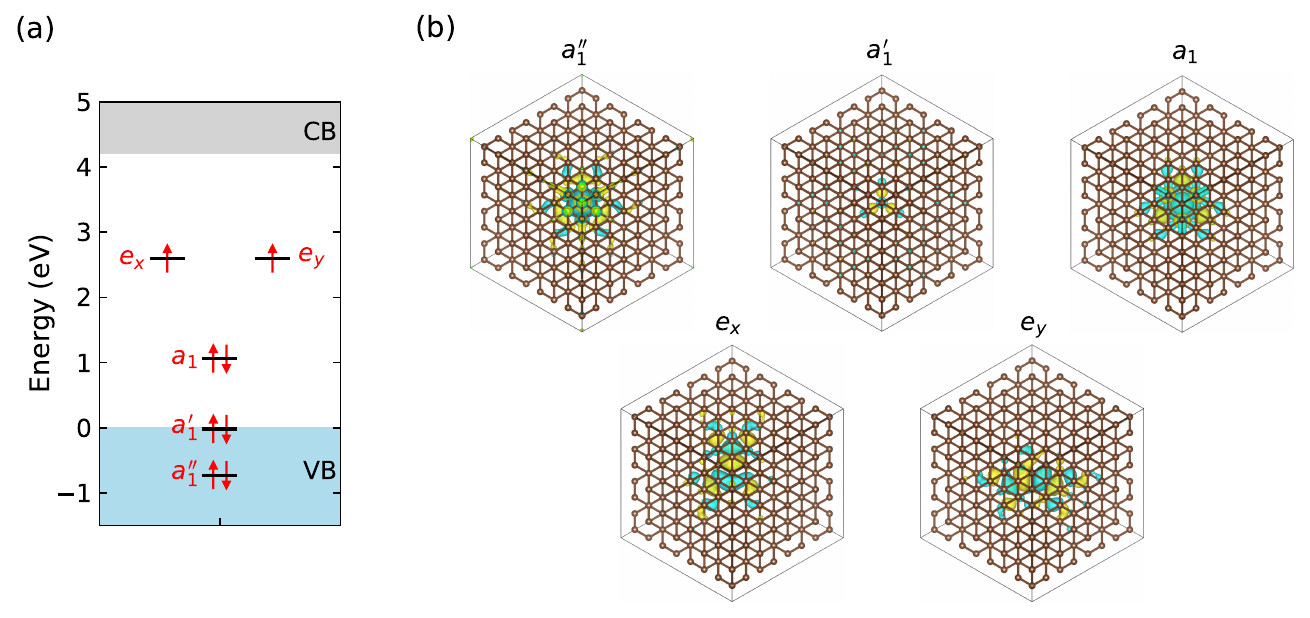}
    \caption{(a) Single-particle defect levels of the NV$^-$ center in diamond computed using spin-restricted DFT/PBE calculations in a 511-atom supercell. (b) Isosurfaces of the squared modulus of Kohn-Sham orbitals that contribute most significantly to the many-body states of the NV$^-$ center. 
    The color (yellow or light blue) indicates the orbital’s sign ($+$ or $-$). The $a_1''$, $a_1'$, $a_1$, $e_x$, and $e_y$ orbitals correspond to Kohn-Sham indices 1009, 1019, 1022, 1023, and 1024 in the 511-atom supercell, respectively.}
    \label{s-fig:ks_orbitals}
\end{figure*}

The many-body wavefunctions of the electronic states of the NV$^-$ center in diamond, determined through group theory analysis using a minimal basis set ($a_1$, $e_x$, and $e_y$ single-particle defect orbitals) and from QDET calculations, are summarized in Table~\ref{s-tab:many-body-wfcs}. The Kohn-Sham (KS) orbitals that are used to build these many-body wavefunctions are illustrated in Fig.~\ref{s-fig:ks_orbitals}. The many-body wavefunctions of the ${}^3\!E$ state after accounting for SOC are presented in Table~\ref{s-tab:many-body-wfcs-3e}. Notably, only the $A_1$ electronic sublevel of the ${}^3\!E$ state has a non-zero SOC matrix element with the ${}^1\!A_1$ state~\cite{maze2011properties,doherty2011negatively}, enabling the intersystem crossing (ISC) denoted here as $\Gamma_{A_1}$. However, the dynamic-Jahn-Teller (DJT) effect introduces vibronic coupling, which mixes the electronic sublevels and facilitates additional ISC processes, including $\Gamma_{E_{1,2}}$ and $\Gamma_{A_2}$. A detailed discussion of these effects is provided in Sec.~\ref{s-sec:djt}.

\renewcommand{\arraystretch}{1.8}
\begin{table*}[ht]
\caption{Many-body wavefunctions of the electronic states of the NV$^-$ center in diamond in the hole representation. Group theory wavefunctions are taken from Ref.~\cite{maze2011properties,doherty2011negatively}, while quantum defect embedding theory (QDET) wavefunctions are calculated using a 511-atom supercell and are converged with respect to the active space selection. The table lists the dominant Slater determinants, with the corresponding orbitals shown in Fig.~\ref{s-fig:ks_orbitals}.}
\label{s-tab:many-body-wfcs}
\centering
\begin{tabularx}{\textwidth}{c|c|c|c}
\hline \hline
State & $S, m_S$ &  Group theory & Quantum defect embedding theory (QDET) \\ \hline
& 1, 1 & $|\overline{e}_x \overline{e}_y\rangle$ & $0.992 |\overline{e}_x \overline{e}_y\rangle$ \\ 
$\left|{}^3\!A_2\right\rangle$ & 1, 0 & $-\frac{1}{\sqrt{2}} |e_y \overline{e}_x\rangle + \frac{1}{\sqrt{2}} |e_x \overline{e}_y\rangle $ & $-0.701 |e_y \overline{e}_x\rangle + 0.701 |e_x \overline{e}_y\rangle$ \\
& 1, $-1$ & $|e_x e_y\rangle$ & $0.992 |e_x e_y\rangle$ \\ \hline
$\left|{}^1\!E_x\right\rangle$  & 0, 0 & $\frac{1}{\sqrt{2}} |e_y \overline{e}_y\rangle - \frac{1}{\sqrt{2}} |e_x \overline{e}_x\rangle$ & $0.669 |e_y \overline{e}_y\rangle - 0.668 |e_x \overline{e}_x\rangle + 0.186 |e_x \overline{a}_1\rangle + 0.186 |a_1 \overline{e}_x\rangle - 0.063 |e_x \overline{a}_1''\rangle -0.063 |a_1''\overline{e}_x\rangle$ \\
$\left|{}^1\!E_y\right\rangle$  & 0, 0 & $\frac{1}{\sqrt{2}} |e_y \overline{e}_x\rangle + \frac{1}{\sqrt{2}} |e_x \overline{e}_y\rangle$ & $0.668 |e_y \overline{e}_x\rangle + 0.186 |e_y \overline{a}_1\rangle + 0.668 |e_x \overline{e}_y\rangle + 0.186 |a_1 \overline{e}_y\rangle - 0.063 |e_y \overline{a}_1''\rangle -0.063 |a_1'' \overline{e}_y \rangle$ \\ \hline
$\left|{}^1\!A_1\right\rangle$  & 0, 0 & $\frac{1}{\sqrt{2}} |e_y \overline{e}_y \rangle + \frac{1}{\sqrt{2}} |e_x \overline{e}_x \rangle$ & $0.687 |e_y \overline{e}_y \rangle + 0.688 |e_x \overline{e}_x \rangle - 0.159 |a_1 \overline{a}_1\rangle + 0.070 |a_1 \overline{a}_1'' \rangle + 0.070 |a_1'' \overline{a}_1\rangle $ \\ \hline
& 1, 1 & $|\overline{e}_x \overline{a}_1\rangle$ & $0.966 |\overline{e}_x \overline{a}_1\rangle - 0.151 |\overline{e}_x \overline{a}_1'\rangle - 0.156 |\overline{e}_x \overline{a}_1''\rangle $ \\
$\left|{}^3\!E_x\right\rangle$ & 1, 0 & $-\frac{1}{\sqrt{2}} |e_x \overline{a}_1\rangle + \frac{1}{\sqrt{2}} |a_1 \overline{e}_x\rangle$ & $- 0.683 |e_x \overline{a}_1\rangle + 0.683 |a_1 \overline{e}_x\rangle + 0.107 |e_x \overline{a}_1'\rangle + 0.110 |e_x \overline{a}_1''\rangle - 0.107 |a_1' \overline{e}_x\rangle - 0.110 |a_1'' \overline{e}_x\rangle$ \\
& 1, $-1$ & $|e_x a_1\rangle$ & $0.966 |e_x a_1\rangle - 0.151 |e_x a_1'\rangle - 0.156 |e_x a_1''\rangle $ \\ \hline
& 1, 1 & $|\overline{e}_y \overline{a}_1\rangle$ & $0.966 |\overline{e}_y \overline{a}_1\rangle - 0.151 |\overline{e}_y \overline{a}_1'\rangle - 0.156 |\overline{e}_y \overline{a}_1''\rangle$ \\
$\left|{}^3\!E_y\right\rangle$ & 1, 0 & $-\frac{1}{\sqrt{2}} |e_y \overline{a}_1\rangle + \frac{1}{\sqrt{2}} |a_1 \overline{e}_y\rangle$ & $- 0.683 |e_y \overline{a}_1\rangle + 0.683 |a_1 \overline{e}_y\rangle + 0.107 |e_y \overline{a}_1'\rangle + 0.110 |e_y \overline{a}_1''\rangle - 0.107 |a_1' \overline{e}_y\rangle - 0.110 |a_1'' \overline{e}_y\rangle$ \\
& 1, $-1$ & $|e_y a_1\rangle$ & $0.966 |e_y a_1\rangle - 0.151 |e_y a_1'\rangle - 0.156 |e_y a_1''\rangle$ \\
\hline \hline
\end{tabularx}
\end{table*}

\renewcommand{\arraystretch}{1.8}
\begin{table*}[ht]
\caption{Many-body wavefunctions of the ${}^3\!E$ state of the NV$^-$ center in diamond inclusive of the spin-orbit interaction~\cite{doherty2011negatively,maze2011properties}. Here, $\left| E_{\pm;1,1} \right\rangle = \mp \frac{1}{\sqrt{2}} \left( \left| E_{x;1,1} \right\rangle \pm i \left| E_{y;1,1} \right\rangle 
 \right)$ and $\left| E_{\pm;1,-1} \right\rangle = \mp \frac{1}{\sqrt{2}} \left( \left| E_{x;1,-1} \right\rangle \pm i \left| E_{y;1,-1} \right\rangle 
 \right)$.}
\label{s-tab:many-body-wfcs-3e}
\centering
\begin{tabular}{c|c}
\hline \hline
State & Wavefunction \\ \hline
$\left| E_1 \right\rangle$ & $\frac{1}{\sqrt{2}} \left[ \frac{1}{\sqrt{2}} \left(\left| E_{x; 1,1} \right\rangle + i \left| E_{y; 1, 1} \right\rangle \right) + \frac{1}{\sqrt{2}} \left( \left| E_{x; 1,-1} \right\rangle - i \left| E_{y; 1,-1} \right\rangle \right) \right] = \frac{1}{\sqrt{2}} \left( - \left| E_{+;1,1} \right \rangle + \left | E_{-;1,-1} \right) \right \rangle$ \\\hline

$\left| E_2 \right\rangle$ & $\frac{1}{\sqrt{2}} \left[ - \frac{1}{\sqrt{2}} \left(\left| E_{x; 1,1} \right\rangle + i \left| E_{y; 1,1} \right\rangle \right) + \frac{1}{\sqrt{2}} \left( \left| E_{x; 1,-1} \right\rangle - i \left| E_{y; 1,-1} \right\rangle \right) \right] = \frac{1}{\sqrt{2}} \left( \left| E_{+;1,1} \right\rangle + \left| E_{-;1,-1} \right\rangle \right) $ \\\hline

$\left| E_x \right\rangle$ & $-\left| E_{y; 1,0} \right\rangle$ \\
$\left| E_y \right\rangle$ & $\left| E_{x; 1,0} \right\rangle$ \\ \hline

$\left| A_2 \right\rangle$ & $\frac{1}{\sqrt{2}} \left[ \left( \frac{1}{\sqrt{2}} \left| E_{x; 1,1} \right\rangle - i \left| E_{y; 1,1} \right\rangle \right) - \frac{1}{\sqrt{2}} \left( \left| E_{x; 1, -1} \right\rangle + i \left| E_{y; 1,-1} \right\rangle \right) \right] = \frac{1}{\sqrt{2}} \left( \left| E_{-;1,1} \right\rangle + \left| E_{+;1,-1} \right\rangle \right)$ \\\hline

$\left| A_1 \right\rangle$ & $\frac{1}{\sqrt{2}} \left[ \frac{1}{\sqrt{2}} \left( \left| E_{x; 1,1} \right\rangle - i \left| E_{y; 1,1} \right\rangle \right) + \frac{1}{\sqrt{2}} \left(\left| E_{x; 1,-1} \right\rangle + i \left| E_{y; 1,-1} \right\rangle \right) \right] = \frac{1}{\sqrt{2}} \left( \left| E_{-;1,1} \right\rangle - \left| E_{+;1,-1} \right\rangle \right)$ \\
\hline \hline
\end{tabular}
\end{table*}

\section{Evaluation of Spin-Orbit Coupling Matrix Elements\label{s-sec:soc}}

\begin{figure*}
    \centering
    \includegraphics[width=12cm]{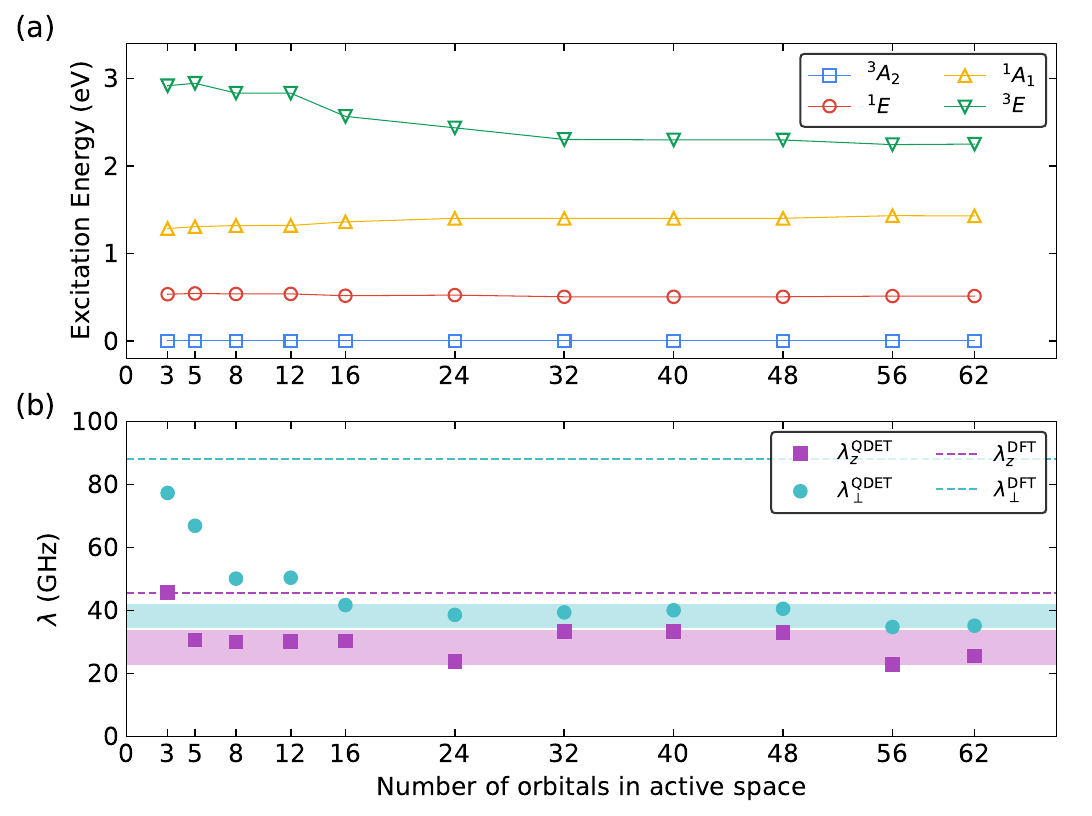}
    \caption{(a) Vertical excitation energies and (b) Spin-orbit coupling (SOC) matrix elements of the many-body states of the NV$^-$ center in diamond, plotted as a function of the number of orbitals in the active space. The results were obtained using quantum defect embedding theory (QDET) in a 511-atom supercell. The SOC matrix elements derived from the many-body wavefunctions obtained via group theory ($\lambda_{z/\perp}^{\mathrm{DFT}}$) are shown as dashed lines.}
    \label{s-fig:qdet_vee_soc}
\end{figure*}

The computed SOC matrix elements, $\lambda_{z/\perp}^{\mathrm{QDET}}$, as a function of the number of orbitals in the QDET active space, are shown in Fig.~\ref{s-fig:qdet_vee_soc}, along with the vertical excitation energies (VEEs) of the many-body states. Notably, $\lambda_{z/\perp}^{\mathrm{QDET}}$ exhibits a convergence behavior similar to that of the VEEs, highlighting that electron correlation effects equally influence the many-body wavefunctions, the SOC matrix elements, and the VEEs.

We investigate the detailed reasons for the decrease in $\lambda_{z/\perp}^{\mathrm{QDET}}$ as the number of orbitals in the QDET active space increases. With only three orbitals in the active space--specifically, the defect orbitals $a_1$, $e_x$, and $e_y$--$\lambda_z^{\mathrm{QDET}}$ remains equal to $\lambda_z^{\mathrm{DFT}}$ because the many-body wavefunctions of ${}^3\!E$ are unchanged relative to those derived through group theory analysis. However, $\lambda_\perp^{\mathrm{QDET}}$ shows a 10.8 GHz reduction compared to $\lambda_\perp^{\mathrm{DFT}}$, due to the mixing of a small portion of the $|a_1\overline{a}_1\rangle$ configuration into the many-body wavefunction of ${}^1\!A_1$; such mixing is not included in the group theory analysis. The inclusion of the $|a_1\overline{a}_1\rangle$ configuration may also explain the underestimation of $\lambda_\perp$ reported in the CASSCF calculations carried out with the C$_{33}$H$_{36}$N$^-$ cluster~\cite{li2024excited}. In Ref.~\cite{li2024excited}, $|a_1\overline{a}_1\rangle$ contributes 21\% to the many-body wavefunction of the ${}^1\!A_1$ state, leading to a significant decrease in $\lambda_\perp$, whereas in our QDET calculations, the contribution is only 3\%.

\begin{figure*}
    \centering
    \includegraphics[width=17cm]{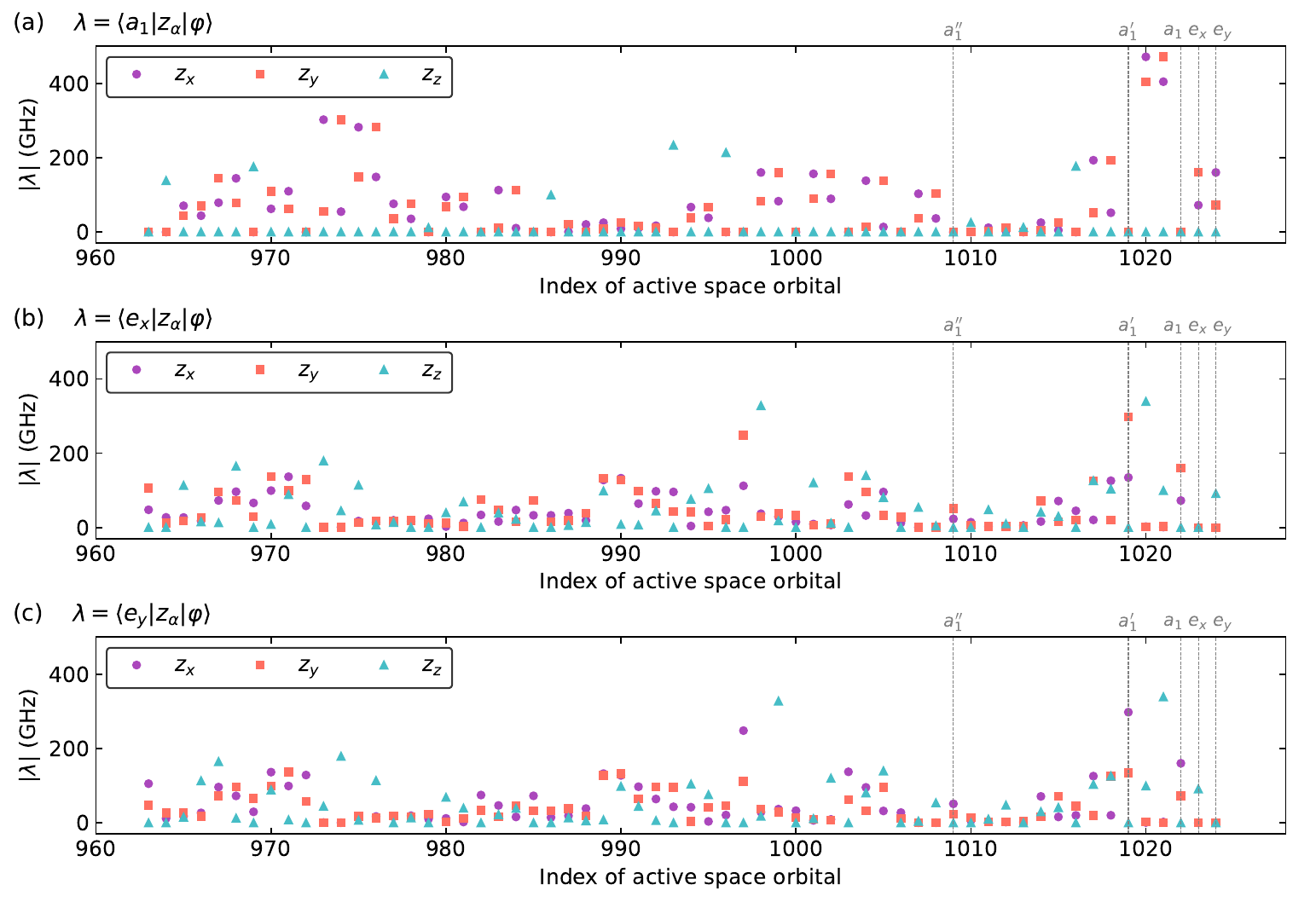}
    \caption{Spin-orbit coupling matrix elements ($\lambda = \langle \varphi_1 | \hat{z}_\alpha | \varphi_2 \rangle$) between defect orbitals and other Kohn-Sham orbitals in the active space, computed in a 511-atom supercell. Panels (a), (b), and (c) correspond to the defect orbitals $a_1$ (orbital 1022), $e_x$ (orbital 1023), and $e_y$ (orbital 1024), respectively. The purple circles, pink squares, and cyan triangles represent the absolute values of the $x$, $y$, and $z$ components of $\lambda$, aligned with the $\langle 1\bar{1}0 \rangle$, $\langle 11\bar{2}\rangle$, and $\langle 111 \rangle$ directions of the diamond lattice, respectively. The defect orbitals $a_1''$ (orbital 1009), $a_1'$ (orbital 1019), $a_1$ (orbital 1022), $e_x$ (orbital 1023), $e_y$ (orbital 1024), shown in Fig.~\ref{s-fig:ks_orbitals}, are indicated as vertical gray dashed lines.}
    \label{s-fig:dft_soc}
\end{figure*}

When the $a_1^\prime$ orbital is included in the active space (leading to more than five orbitals in the active space), $\lambda_\perp^{\mathrm{QDET}}$ decreases significantly. This reduction is due to the large SOC matrix elements between the $e_{x/y}$ and $a_1^\prime$ orbitals, as illustrated in Fig.~\ref{s-fig:dft_soc}(b) and (c). A further decrease in $\lambda_\perp^{\mathrm{QDET}}$ is observed when the $a_1^{\prime\prime}$ orbital is included in the active space, leading to sixteen or more orbitals in the active space in Fig.~\ref{s-fig:qdet_vee_soc}). The decrease is caused by the non-negligible SOC matrix elements between the $e_{x/y}$ and $a_1^{\prime\prime}$ orbitals. The results obtained by including $a_1^{\prime}$ and $a_1^{\prime\prime}$ also explains the decrease in the VEE of the ${}^3\!E$ state. Although this analysis focuses solely on the configurations that contribute most significantly to the many-body wavefunctions, it highlights the critical role of electron correlation effects in accurately computing SOC matrix elements.

\renewcommand{\arraystretch}{1.5}
\begin{table*}[ht]
\caption{Spin-orbit coupling (SOC) matrix elements computed in this work, compared to values reported in previous theoretical and experimental studies. All values are given in GHz.}
\label{s-tab:soc}
\centering
\setlength{\tabcolsep}{8pt}
\begin{tabularx}{\textwidth}{c|c|c|c}
\hline\hline
& $\lambda_z$ & $p\lambda_z$ \textsuperscript{a} & $\lambda_\perp$ \\ \hline
\hspace{12mm} DFT (PBE, ONCV, this work), 511-atom supercell \hspace{12mm} & 45.7 & 13.9 & 88.1 \\
DFT (PBE, ONCV, this work), dilute limit & 37.5 & 11.4 & 70.6 \\
QDET (This work), 511-atom supercell & $28.1\pm 5.2$ & $8.5 \pm 1.6$ & $36.6 \pm 3.5$ \\
QDET (This work), dilute limit & $23.1 \pm 4.3$ & $7.0 \pm 1.3$ & $29.4 \pm 2.8$ \\
DFT (PBE, PAW), dilute limit~\cite{thiering2017ab} \textsuperscript{b} & 20.4 & 6.20 & \\
DFT (HSE, PAW), dilute limit~\cite{thiering2017ab} \textsuperscript{b} & 15.78 & 4.80 & 56.32 \\
CASSCF, C$_{33}$H$_{36}$N$^-$ cluster~\cite{li2024excited} & 14.21 & 4.32 & 3.96 \\
CI-cRPA, 511-atom supercell~\cite{neubauer2024spin} & 20.6 & 6.3 & 59.3 \textsuperscript{c} \\
Expt.~\cite{batalov2009low,bassett2014ultrafast,goldman2015state,goldman2017erratum} & $17.5\pm0.1$ \textsuperscript{d} & $5.33\pm0.03$ & $21.1\pm3.6$ \textsuperscript{e} \\
 \hline\hline
\end{tabularx}

\vspace{1mm}
\begin{minipage}{\textwidth}
\raggedright
\hspace{0mm}\textsuperscript{a} $p$ is the Ham reduction factor from the dynamic Jahn-Teller effects of the ${}^3\!E$ state. $p = 0.304$~\cite{thiering2017ab} is used in this work.

\hspace{0mm}\textsuperscript{b} SOC matrix elements are computed as $\lambda_z = \frac{1}{2} \langle e_+ | \hat{H}_{\mathrm{SO}} | e_+ \rangle $ and $\lambda_\perp = \frac{1}{\sqrt{2}} \langle e_+^\downarrow | \hat{H}_{\mathrm{SO}} | a_1^{\uparrow} \rangle$, where $|e_\pm\rangle = \frac{1}{\sqrt{2}} \left(|e_x\rangle \pm i |e_y\rangle \right)$~\cite{thiering2017ab}.

\hspace{0mm}\textsuperscript{c} $\lambda_\perp$ is computed as $\dfrac{1}{\sqrt{2}} \left( \left|\left \langle {}^1\!A_1 \right| \hat{H}_{\mathrm{SO}} \left| 
{}^3\!E_{x; 1, 1} \right \rangle\right|^2 + \left|\left \langle {}^1\!A_1 \right| \hat{H}_{\mathrm{SO}} \left| 
{}^3\!E_{x; 1, -1} \right \rangle\right|^2 + \left|\left \langle {}^1\!A_1 \right| \hat{H}_{\mathrm{SO}} \left| 
{}^3\!E_{y; 1, 1} \right \rangle\right|^2 + \left|\left \langle {}^1\!A_1 \right| \hat{H}_{\mathrm{SO}} \left| 
{}^3\!E_{y; 1, -1} \right \rangle\right|^2\right)^{1/2} = 59.3$ GHz, where $\left|\left \langle {}^1\!A_1 \right| \hat{H}_{\mathrm{SO}} \left| 
{}^3\!E_{x; 1, 1} \right \rangle\right| = \left|\left \langle {}^1\!A_1 \right| \hat{H}_{\mathrm{SO}} \left| 
{}^3\!E_{x; 1, -1} \right \rangle\right| = \left|\left \langle {}^1\!A_1 \right| \hat{H}_{\mathrm{SO}} \left| 
{}^3\!E_{y; 1, 1} \right \rangle\right| =  \left|\left \langle {}^1\!A_1 \right| \hat{H}_{\mathrm{SO}} \left| 
{}^3\!E_{y; 1, -1} \right \rangle\right| = 0.1735\ {\mathrm{meV}} = 41.95$ GHz~\cite{neubauer2024spin}.

\hspace{0mm}\textsuperscript{d} $\lambda_z = 17.53 \pm 0.10$ GHz is obtained using the experimentally derived $p\lambda_z = 5.33 \pm 0.03$ GHz~\cite{batalov2009low} together with $p = 0.304$~\cite{thiering2017ab}.

\hspace{0mm}\textsuperscript{e} $\lambda_\perp = 21.06 \pm 3.62$ GHz is obtained using the approximated relation $\lambda_\perp = (1.2 \pm 0.2) \lambda_z$~\cite{goldman2015state,goldman2017erratum}.
\end{minipage}
\end{table*}

\section{Role of the Dynamic Jahn-Teller Effect in Intersystem Crossing Rates\label{s-sec:djt}}

At temperatures near 0 K, the ISC rate, $\Gamma_{A_1}$, is calculated using Fermi's golden rule:
\begin{equation}
    \Gamma_{A_1} = \frac{2\pi}{\hbar} \sum_n \left|\left\langle \widetilde{A}_1  \right| \hat{H}_{\mathrm{SO}} \left| {}^1\!\widetilde{A}_{1,n}\right\rangle \right|^2 \delta \left(E_{\widetilde{A}_1} - E_{{}^1\!\widetilde{A}_{1,n}} \right)
    \label{s-eq:isc_rate_0}
\end{equation}
where $\widetilde{A}_1$ represents the first coupled electronic-vibrational (vibronic) level of the ${}^3\!E$ state with $A_1$ symmetry and energy $E_{\widetilde{A}_1}$. ${}^1\!\widetilde{A}_{1,n}$ refers to the $n$th vibronic level of the ${}^1\!A_1$ state, with energy given by $E_{{}^1\!\widetilde{A}_{1,n}} = E_{{}^1\!\widetilde{A}_{1,0}} + \widetilde{E}_{{}^1\!\widetilde{A}_{1,n}}$, where $\widetilde{E}_{{}^1\!\widetilde{A}_{1,n}}$ denotes the energy relative to the lowest vibronic level, ${}^1\!\widetilde{A}_{1,0}$, which has energy $E_{{}^1\!\widetilde{A}_{1,0}}$. $\hat{H}_{\mathrm{SO}}$ is the many-body SOC operator.

The vibronic level can be expressed as a direct product of the vibronic wavefunction associated with the $e$-type phonon modes and the vibrational wavefunctions associated with the $a_1$-type phonon modes,
\begin{equation}
    \left| \widetilde{A}_1 \right\rangle = \left| \widetilde{A}_1^e \right\rangle \otimes \left| \chi^{a_1} \right\rangle,
\end{equation}
and
\begin{equation}
    \left| {}^1\!\widetilde{A}_{1,n} \right\rangle = \left|{}^1\!\widetilde{A}^{e}_{1,n_e} \right\rangle \otimes \left| \chi^{a_1}_{n_{a_1}} \right\rangle,
\end{equation}
and the relative energy of the vibronic level can be written as a sum of contributions from the $e$-type and $a_1$-type phonon modes:
\begin{equation}
    \widetilde{E}_{{}^1\!\widetilde{A}_{1,n}} = \widetilde{E}_{{}^1\!\widetilde{A}^e_{1,n_e}} + \widetilde{E}_{\chi^{a_1}_{n_{a_1}}}.
\end{equation}
Substituting these expression into Eq.~\ref{s-eq:isc_rate_0}, we obtain:
\begin{equation}
\begin{aligned}
    \Gamma_{A_1} &= \frac{2\pi}{\hbar} \sum_{n_e} \sum_{n_{a_1}} \left| \left \langle \chi^{a_1} \right| \otimes \left\langle \widetilde{A}_1^e  \right| \hat{H}_{\mathrm{SO}} \left| {}^1\!\widetilde{A}^e_{1,n_e}\right\rangle \otimes \left| \chi_{n_{a_1}}^{a_1} \right \rangle \right|^2 \delta \left(\Delta - \widetilde{E}_{{}^1\!\widetilde{A}^e_{1,n_e}} - \widetilde{E}_{\chi^{a_1}_{n_{a_1}}} \right),
\end{aligned}
\end{equation}
where $\Delta = E_{\widetilde{A}_1} -{E}_{{}^1\!\widetilde{A}_{1,0}}$ represents the energy gap between the lowest vibronic levels of ${}^3\!E$ and ${}^1\!A_1$. The contribution of $a_1$-type phonon modes is accounted for by the VOF:
\begin{equation}
    F^{a_1}(\varepsilon) = \sum_{n_{a_1}} \left| \left \langle \chi^{a_1} \right. \left| \chi_{n_{a_1}}^{a_1} \right \rangle \right|^2 \delta \left(\varepsilon -  \widetilde{E}_{\chi^{a_1}_{n_{a_1}}} \right),
    \label{s-eq:spectral_fxn_a1}
\end{equation}
which effectively integrates out the $a_1$-type phonon mode contributions. Substituting $F^{a_1}(\varepsilon)$ into the ISC rate expression yields:
\begin{equation}
    \Gamma_{A_1} = \frac{2\pi}{\hbar} \sum_{n_e} \left| \left\langle \widetilde{A}_1^e  \right| \hat{H}_{\mathrm{SO}} \left| {}^1\!\widetilde{A}^e_{1,n_e}\right\rangle \right|^2 F^{a_1} \left(\Delta - \widetilde{E}_{{}^1\!\widetilde{A}^e_{1,n_e}} \right).
    \label{s-eq:isc_rate_djt}
\end{equation}
This formulation separates the effects of the $e$-type and $a_1$-type phonon modes, simplifying the computation of the ISC rate by addressing each contribution independently.

The evaluation of the ISC rate via Eq.~\ref{s-eq:isc_rate_djt} requires addressing the DJT effect for the $e$-type phonon modes and the ${}^3\!E$ state. The DJT effect couples the $A_1$, $A_2$, $E_1$, and $E_2$ electronic sublevels of the ${}^3\!E$ state through vibronic interactions, leading to the following expressions for their vibronic wavefunctions~\cite{thiering2017ab}:
\begin{equation}
    \left|\widetilde{A}_1^e \right\rangle = \sum_i [c_i^{\widetilde{A}_1^e} \left |A_1\right\rangle \otimes \left|\chi_i^e \left(A_1\right)\right\rangle + \frac{d_i^{\widetilde{A}_1^e}}{\sqrt{2}}(\overbrace{\left|E_1\right\rangle \otimes \left|\chi^e_i \left(E_1\right)\right\rangle + \left|E_2\right\rangle \otimes \left|\chi^e_i \left(E_2\right)\right\rangle}^{\subset A_1}) + f_i^{\widetilde{A}_1^e} \overbrace{\left|A_2\right\rangle \otimes \left|\chi_i^e \left(A_2\right)\right\rangle}^{\subset A_1}],
\end{equation}
\begin{equation}
    \left|\widetilde{A}_2^e\right\rangle = \sum_i[c_i^{\widetilde{A}_2^e} \left|A_2\right\rangle \otimes \left|\chi^e_i\left(A_1\right)\right\rangle + \frac{d_i^{\widetilde{A}_2^e}}{\sqrt{2}}(\overbrace{\left|E_1\right\rangle \otimes \left|\chi_i^e\left(E_2\right)\right\rangle - \left|E_2\right\rangle \otimes \left|\chi_i^e\left(E_1\right)\right\rangle}^{\subset A_2}) + f_i^{\widetilde{A}_2^e} \overbrace{\left|A_1\right\rangle \otimes \left|\chi_i^e\left(A_2\right)\right\rangle}^{\subset A_2}],
\end{equation}
\begin{equation}
\begin{aligned}
    \left|\widetilde{E}_1^e\right\rangle = \sum_i[ &c_i^{\widetilde{E}_1^e} \left|E_1\right\rangle \otimes \left|\chi^e_i\left(A_1\right)\right\rangle + \frac{d_i^{\widetilde{E}_1^e}}{\sqrt{2}}(\overbrace{\left|A_1\right\rangle \otimes \left|\chi_i^e\left(E_1\right)\right\rangle + \left|A_2\right\rangle \otimes \left|\chi^e_i\left(E_2\right)\right\rangle}^{\subset E_1}) + f_i^{\widetilde{E}_1^e} \overbrace{\left|E_2\right\rangle \otimes \left|\chi_i^e\left(A_2\right)\right\rangle}^{\subset E_1}\\
    &+ \dfrac{g_i^{\widetilde{E}_1^e}}{\sqrt{2}} (\overbrace{\left|E_1\right\rangle \otimes \left|\chi_i^e\left(E_1\right)\right\rangle - \left|E_2\right\rangle \otimes \left|\chi^e_i\left(E_2\right)\right\rangle}^{\subset E_1})],
\end{aligned}
\end{equation}
\begin{equation}
\begin{aligned}
    \left|\widetilde{E}_2^e \right\rangle =\sum_i[&c_i^{\widetilde{E}_2^e} \left|E_2\right\rangle \otimes \left|\chi^e_i\left(A_1\right)\right\rangle + \frac{d_i^{\widetilde{E}_2^e}}{\sqrt{2}}(\overbrace{\left|A_1\right\rangle \otimes \left|\chi^e_i\left(E_2\right)\right\rangle-\left|A_2\right\rangle \otimes \left|\chi^e_i\left(E_1\right)\right\rangle}^{\subset E_2}) + f_i^{\widetilde{E}_2^e} \overbrace{\left|E_1\right\rangle \otimes \left|\chi^e_i\left(A_2\right)\right\rangle}^{\subset E_2}\\
    &+ \dfrac{g_i^{\widetilde{E}_2^e}}{\sqrt{2}} (\overbrace{\left|E_1\right\rangle \otimes \left|\chi_i^e\left(E_2\right)\right\rangle + \left|E_2\right\rangle \otimes \left|\chi^e_i\left(E_1\right)\right\rangle}^{\subset E_2})].
\end{aligned}
\end{equation}
Here, $\chi_i^e(A_1)$, $\chi_i^e(A_2)$, $\chi_i^e(E_1)$, and $\chi_i^e(E_2)$ are the $i$th symmetry-adapted vibration wavefunctions for each irreducible representation. Due to DJT mixing, the $|A_1\rangle$ electronic sublevel contributes to the $\widetilde{A}_2^e$, $\widetilde{E}_1^e$, and $\widetilde{E}_2^e$ vibronic levels. Consequently, these levels also participate in ISC to the singlet ${}^1\!A_1$. The vibronic level $\left| {}^1\!\widetilde{A}^e_{1,n_e}\right\rangle$, neglecting the pseudo-Jahn-Teller effect with the ${}^1\!E$ state (which only slightly perturbs the ${}^1\!A_1$ vibronic levels~\cite{thiering2018theory,jin2022vibrationally}), can be expressed as:
\begin{equation}
    \left| {}^1\!\widetilde{A}^e_{1,n_e}\right\rangle = \left| {}^1\!A_1 \right\rangle \otimes | \chi_{n_e}^e \rangle.
\end{equation}
With these considerations in mind, the ISC rate for the $\widetilde{A}_1$ vibronic level is written as:
\begin{equation}
    \Gamma_{A_1} = \frac{2\pi}{\hbar} \left| \left\langle A_1  \right| \hat{H}_{\mathrm{SO}} \left| {}^1\!A_{1}\right\rangle \right|^2 \sum_{i} \left|c_i^{\tilde{A}_1^e} \right|^2  F^{a_1} \left(\Delta - \widetilde{E}_{\chi^e_i} \right) = \frac{4\pi}{\hbar} |\lambda_\perp|^2 F_{A_1}(\Delta),
    \label{s-eq:isc_rate_gamma_a1_final}
\end{equation}
where $F_{A_1}(\Delta) = \sum_{i} \left|c_i^{\tilde{A}_1^e}\right|^2 F^{a_1} \left(\Delta - \widetilde{E}_{\chi^e_i} \right)$ accounts for the DJT effect. Here, we used the relation $\left\langle \chi_i^e | \chi_{n_e}^e\right\rangle = \delta_{in_e}$ and $\left\langle E_1 \right| \hat{H}_{\mathrm{SO}} \left| {}^1\!A_1 \right\rangle = \left\langle E_2 \right| \hat{H}_{\mathrm{SO}} \left| {}^1\!A_1 \right\rangle = \left\langle A_2 \right| \hat{H}_{\mathrm{SO}} \left| {}^1\!A_1 \right\rangle = 0$. Similarly, the ISC rates for the $\widetilde{E}_{1,2}$ and $\widetilde{A}_2$ vibronic levels are given by:
\begin{equation}
    \Gamma_{E_{1,2}} = \frac{2\pi}{\hbar} \left| \left\langle A_1  \right| \hat{H}_{\mathrm{SO}} \left| {}^1\!A_{1}\right\rangle \right|^2 \sum_{i} \frac{\Big|d_i^{\tilde{E}_{1,2}^e}\Big|^2}{2} F^{a_1} \left(\Delta - \widetilde{E}_{\chi^e_i} \right) = \frac{4\pi}{\hbar} |\lambda_\perp|^2 F_{E_{1,2}}(\Delta),
    \label{s-eq:isc_rate_gamma_e12_final}
\end{equation}
\begin{equation}
    \Gamma_{A_{2}} = \frac{2\pi}{\hbar} \left| \left\langle A_1  \right| \hat{H}_{\mathrm{SO}} \left| {}^1\!A_{1}\right\rangle \right|^2 \sum_{i} \left|f_i^{\tilde{A}_2^e}\right|^2 F^{a_1} \left(\Delta - \widetilde{E}_{\chi^e_i} \right) = \frac{4\pi}{\hbar} |\lambda_\perp|^2 F_{A_{2}}(\Delta),
    \label{s-eq:isc_rate_gamma_a2_final}
\end{equation}
where $F_{E_{1,2}}(\Delta) = \sum_{i} \dfrac{\big|d_i^{\tilde{E}_{1,2}^e}\big|^2}{2} F^{a_1} \left(\Delta - \widetilde{E}_{\chi^e_i} \right)$ and $F_{A_2}(\Delta) = \sum_{i} \left|f_i^{\tilde{A}_2^e}\right|^2  F^{a_1} \left(\Delta - \widetilde{E}_{\chi^e_i} \right)$.

To simplify the treatment of the DJT effect, we adopted the two effective phonon mode approximation as described in Ref.~\cite{thiering2017ab}. Using the electronic basis set $\left[{}^3\!E_{x;1,-1}, {}^3\!E_{y;1,-1}, {}^3\!E_{x;1,1}, {}^3\!E_{y;1,1}\right]$, the effective Hamiltonian is written as:
\begin{equation}
    \hat{H}_{\mathrm{eff}} = \lambda_z \left( \begin{matrix}
        \sigma_y & 0 \\
        0 & - \sigma_y \\
    \end{matrix} \right) + \left( \begin{matrix}
        \hat{H}_{\mathrm{DJT}} & 0 \\
        0 & \hat{H}_{\mathrm{DJT}} \\
    \end{matrix} \right),
    \label{s-eq:eff_h}
\end{equation}
where:
\begin{equation}
    \hat{H}_{\mathrm{DJT}} = \hbar \omega_{e}\left(\hat{a}_{x}^{\dagger} \hat{a}_{x} + \hat{a}_{y}^{\dagger} \hat{a}_{y} + 1 \right) + F\left(\hat{Q}_x \sigma_{z}-\hat{Q}_y \sigma_{x}\right) + G \left[\left(\hat{Q}_x^{2}-\hat{Q}_y^{2}\right) \sigma_{z}+2 \hat{Q}_x \hat{Q}_y \sigma_{x}\right].
\end{equation}
Here, $\sigma_x$, $\sigma_y$, and $\sigma_z$ are the Pauli matrices. $\hat{Q}_{x/y} = \frac{1}{\sqrt{2}} \left( 
\hat{a}_{x/y}^{\dagger} + \hat{a}_{x/y} \right)$ are unitless position operators. $\hbar\omega_e$, $F$, and $G$ are the effective phonon energy, linear DJT coupling constant, and quadratic DJT coupling constant, respectively. The DJT parameters are derived by fitting the adiabatic potential energy surface (APES) of the ${}^3\!E$ state. In this work, we use $\hbar\omega_e = 77.6$ meV, $F = 75.46$ meV, and $G = 4.74$ meV as derived using the APES from Ref.~\cite{thiering2017ab}. The first four vibronic levels, obtained by solving the effective Hamiltonian Eq.~\ref{s-eq:eff_h}, form two degenerate groups and are expressed as follows:
\begin{equation}
    \left| \widetilde{\Phi}_{-;1,-1}^e \right\rangle = \sum_{nm} \left( c_{n m} \left| {}^3\!E_{-;1,-1} \right\rangle \otimes |n, m\rangle + d_{n m} \left| {}^3\!E_{+;1,-1} \right \rangle \otimes |n, m\rangle \right),
    \label{s-eq:vib_minus_minus1}
\end{equation}
\begin{equation}
    \left| \widetilde{\Phi}_{+;1,1}^e \right\rangle = \sum_{nm} \left( c^{*}_{n m} \left| {}^3\!E_{+;1,1} \right\rangle \otimes |n, m\rangle + d^{*}_{n m} \left| {}^3\!E_{-;1,1} \right \rangle \otimes |n, m\rangle \right),
    \label{s-eq:vib_plus_plus1}
\end{equation}
and
\begin{equation}
    \left| \widetilde{\Phi}_{+;1,-1}^e \right\rangle = \sum_{nm} \left( c^{\prime*}_{n m} \left| {}^3\!E_{+;1,-1} \right\rangle \otimes |n, m\rangle + d^{\prime*}_{n m} \left| {}^3\!E_{-;1,-1} \right \rangle \otimes |n, m\rangle \right),
    \label{s-eq:vib_plus_minu1}
\end{equation}
\begin{equation}
    \left| \widetilde{\Phi}_{-;1,1}^e \right\rangle = \sum_{nm} \left( c_{n m}^{\prime} \left| {}^3\!E_{-;1,1} \right\rangle \otimes |n, m\rangle + d_{n m}^{\prime} \left| {}^3\!E_{+;1,1} \right \rangle \otimes |n, m\rangle \right),
    \label{s-eq:vib_minus_plus1}
\end{equation}
where $\left|{}^3\!E_{\pm}\right\rangle = \mp \frac{1}{\sqrt{2}} \left(\left|{}^3\!E_{x}\right\rangle \pm i \left|{}^3\!E_{y}\right\rangle \right)$, and $|n, m\rangle$ are the two-dimensional harmonic oscillator basis states. Expansions with $m + n \leq 5$ are sufficient to converge the results. The vibronic levels $\left|\widetilde{A}_1^e\right\rangle$, $\left|\widetilde{A}_2^e\right\rangle$, $\left|\widetilde{E}_1^e\right\rangle$, and $\left|\widetilde{E}_2^e\right\rangle$ are given by:
\begin{equation}
\begin{aligned}
    \left| \widetilde{A}_1^e \right \rangle &= \frac{1}{\sqrt{2}} \left( \left| \widetilde{\Phi}_{-;1,1}^e \right\rangle - \left| \widetilde{\Phi}_{+;1,-1}^e \right\rangle \right) \\
    \left| \widetilde{A}_2^e \right \rangle &= \frac{1}{\sqrt{2}} \left( \left| \widetilde{\Phi}_{-;1,1}^e \right\rangle + \left| \widetilde{\Phi}_{+;1,-1}^e \right\rangle \right) \\
    \left| \widetilde{E}_1^e \right \rangle &= \frac{1}{\sqrt{2}} \left( \left| \widetilde{\Phi}_{-;1,-1}^e \right\rangle - \left| \widetilde{\Phi}_{+;1,1}^e \right\rangle \right) \\
    \left| \widetilde{E}_2^e \right \rangle &= \frac{1}{\sqrt{2}} \left( \left| \widetilde{\Phi}_{-;1,-1}^e \right\rangle + \left| \widetilde{\Phi}_{+;1,1}^e \right\rangle \right). \\
    \label{s-eq:four_vib_levels}
\end{aligned}
\end{equation}
The coefficients $c_i^{\tilde{A}_1^e}$, $d_i^{\tilde{E}_{1,2}^e}$, and $f_i^{\tilde{A}_2^e}$ used in Eqs.~\ref{s-eq:isc_rate_gamma_a1_final}--\ref{s-eq:isc_rate_gamma_a2_final} are obtained by transforming Eq.~\ref{s-eq:four_vib_levels} using the electronic wavefunctions in Table~\ref{s-tab:many-body-wfcs-3e} and symmetry-adapted vibrational wavefunctions~\cite{thiering2017ab}, and the results are shown in Table~\ref{s-tab:vib_coeff}. The magnitudes of the coefficients $\left|c_i^{\tilde{A}_1^e} \right|^2$ and $\left|d_i^{\tilde{E}_{1,2}^e} \right|^2$ are significant, resulting in observable ISC rates $\Gamma_{A_1}$ and $\Gamma_{E_{1,2}}$. In contrast, the coefficients $\left|f_i^{\tilde{A}_2^e} \right|^2$ are nearly negligible, leading to an insignificant ISC rate $\Gamma_{A_2}$, which is therefore not considered in this work. The energy $\widetilde{E}_{\chi_i^e}$, used in the calculation of the VOFs $F_{A_1}(\Delta)$ and $F_{E_{1,2}}(\Delta)$, is determined as $n_i \hbar \omega_e$.

\renewcommand{\arraystretch}{1.8}
\begin{table*}[ht]
\caption{Coefficients of vibronic levels for the calculations of vibrational overlap functions (VOFs) in Eqs.~\ref{s-eq:isc_rate_gamma_a1_final}--\ref{s-eq:isc_rate_gamma_a2_final}.}
\label{s-tab:vib_coeff}
\centering
\begin{tabular}{c|cc|cc|cc}
\hline \hline
$\quad i\quad$ & $\quad \left|c_i^{\tilde{A}_1^e} \right|^2 \quad$ & $\quad n_i \quad$ & $\quad \left|d_i^{\tilde{E}_{1,2}^e} \right|^2 \quad$ & $\quad n_i \quad$ & $\quad \left|f_i^{\tilde{A}_2^e} \right|^2 \quad$ & $\quad n_i \quad$ \\ \hline
1 & 0.5751 & 0 & 0.3303 & 1 & $< 0.0001$ & 3 \\
2 & 0.0712 & 2 & 0.0039 & 2 & $\dots$ & $\dots$ \\
3 & 0.0026 & 3 & 0.0133 & 3 & & \\
4 & 0.0016 & 4 & 0.0007 & 4 & & \\
5 & $\dots$ & $\dots$ & $\dots$ & $\dots$ & & \\
\hline \hline
\end{tabular}
\end{table*}

\section{Computation of Vibrational Overlap Functions\label{s-sec:spectral_fxn}}

Following Sec.~\ref{s-sec:djt}, we compute the VOF for the ${}^3\!E \to {}^1\!A_1$ ISC transition involving the $a_1$-type phonon modes using the HR theory and the generating function approach~\cite{huang1950theory,alkauskas2014first,jin2021photoluminescence}. The equilibrium atomic geometries of the ${}^3\!E$ and ${}^1\!A_1$ states were obtained using spin-conserving and spin-flip TDDFT, respectively, while the phonon modes of the ${}^1\!A_1$ state were calculated with the frozen phonon approach, using analytical nuclear forces from spin-flip TDDFT. The results are extrapolated to the dilute limit, approximated by a 13823-atom supercell.

\begin{figure*}
    \centering
    \includegraphics[width=16cm]{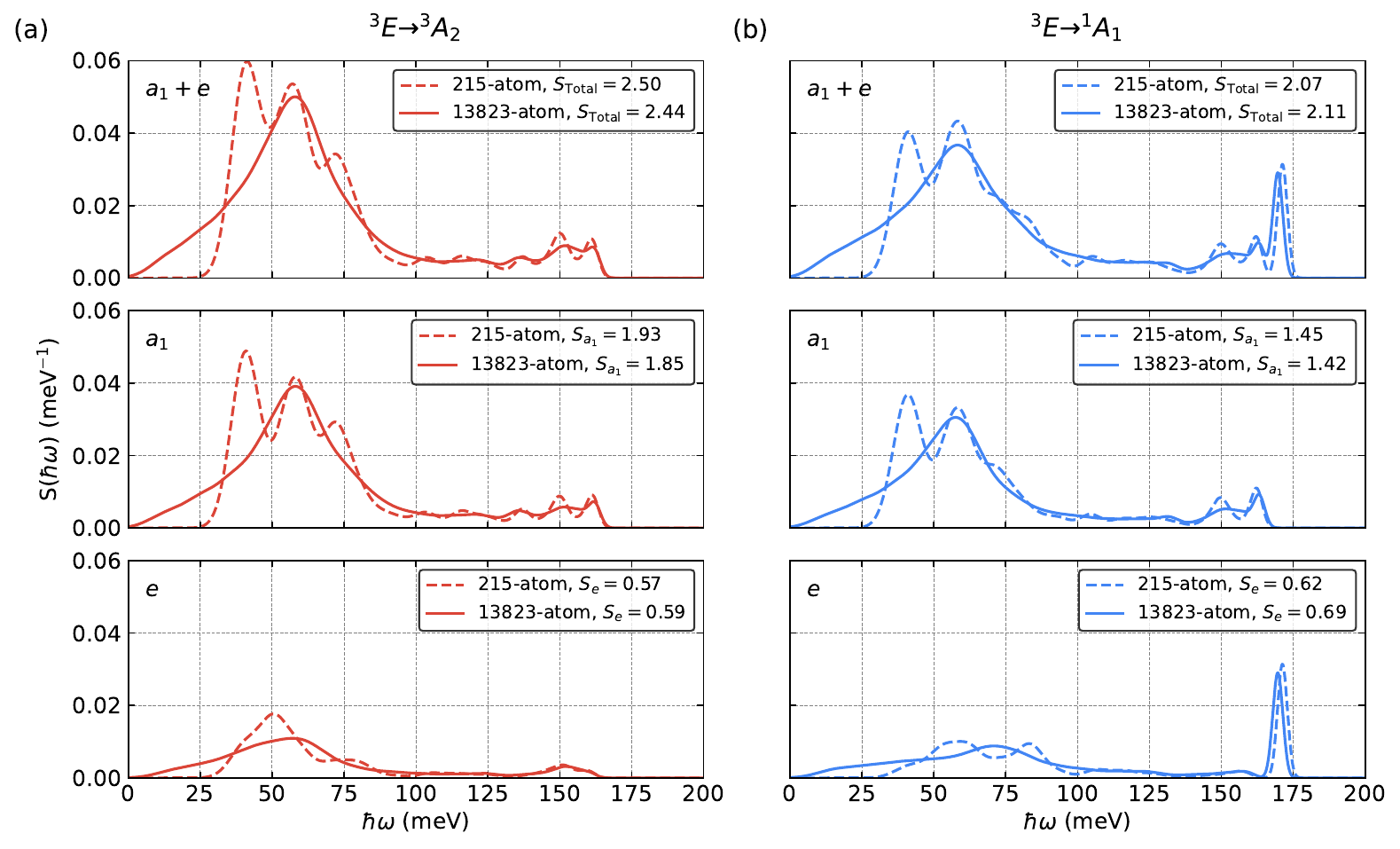}
    \caption{Computed spectral densities of electron-phonon interaction (distributions of partial Huang-Rhys factors) for the (a) ${}^3\!E \to {}^3\!A_2$ transition and (b) ${}^3\!E \to {}^1\!A_1$ transition. The upper, middle, and bottom panels display the total spectral density ($a_1 + e$), the contribution of $a_1$-type phonon modes, and the contribution of $e$-type phonon modes, respectively. Spectral densities computed using the 215-atom supercell are shown as dashed lines, while the extrapolated results to the 13823-atom supercell are shown as solid lines. The total Huang-Rhys factors ($S_{\mathrm{Total}}$, $S_{a_1}$, and $S_e$) are provided in the inset.}
    \label{s-fig:spectral_densities}
\end{figure*}

The computed spectral densities, representing the distributions of partial HR factors as a function of the phonon energy, are shown in Fig.~\ref{s-fig:spectral_densities}. For comparison, the spectral densities for the ${}^3\!E \to {}^3\!A_2$ photoluminescence (PL) transition are also included. Notably, finite-size effects significantly impact the spectral densities by smoothing out artificial peaks and incorporating contributions from low-frequency acoustic modes. The general shapes of the spectral densities for the ${}^3\!E \to {}^3\!A_2$ and ${}^3\!E \to {}^1\!A_1$ transitions are similar due to their comparable electronic configurations ($a_1^2e_x^1e_y^1$). However, the intensity of the ${}^3\!E \to {}^3\!A_2$ transition is 16\% larger than that of the ${}^3\!E \to {}^1\!A_1$ transition (30\% when considering only the $a_1$-type phonon modes). This difference originates from electron correlation effects on the ${}^1\!A_1$ state and is critical for accurately calculating ISC rates. Previous studies considered the VOF of the ${}^3\!E \to {}^3\!A_2$ transition, including both $a_1$- and $e$-type phonon mode contributions~\cite{goldman2015state,li2024excited}, leading to inaccuracies in the calculated ISC rates.

\begin{figure*}
    \centering
    \includegraphics[width=16cm]{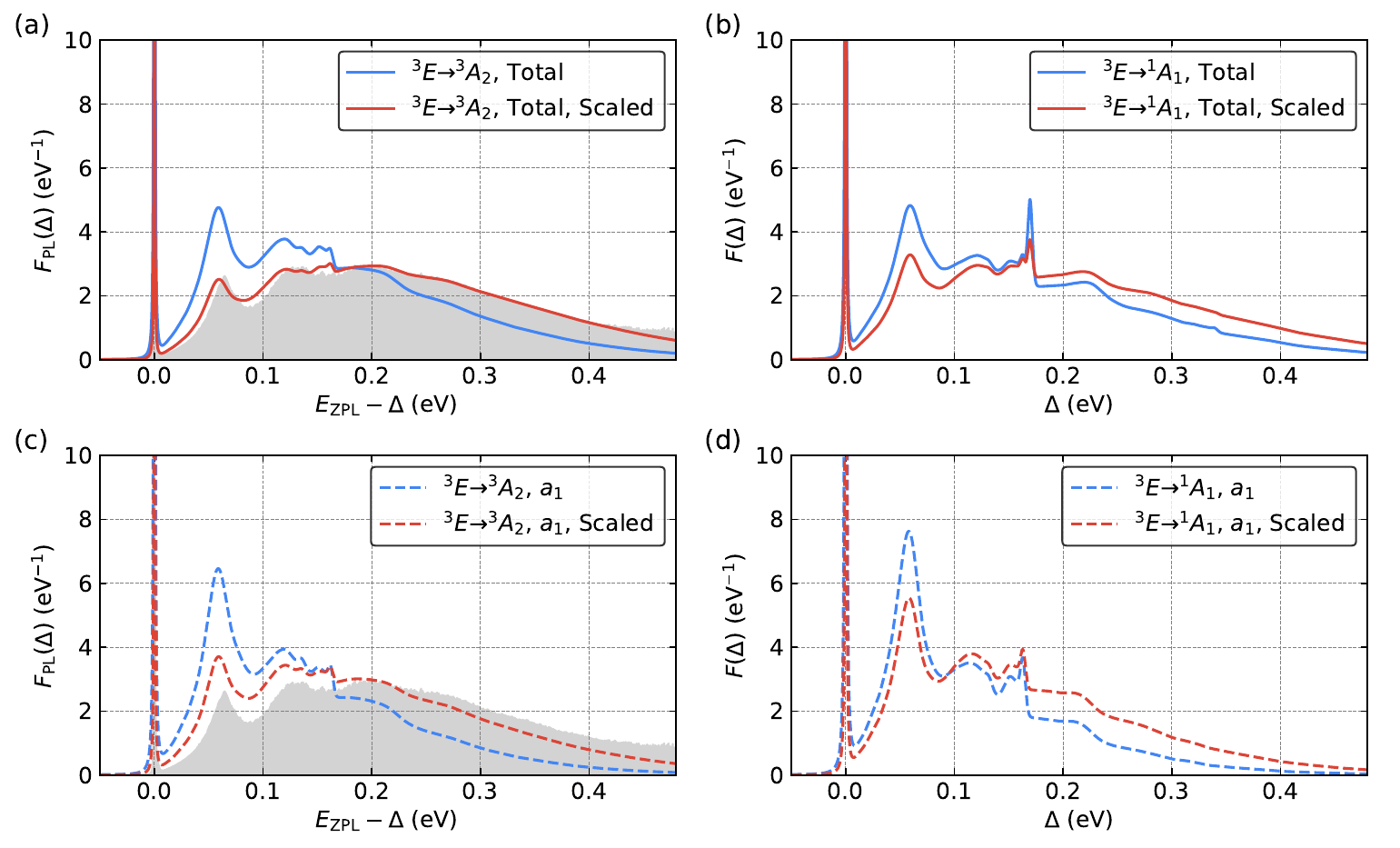}
    \caption{Photoluminescence (PL) spectral functions for the ${}^3\!E \to {}^3\!A_2$ transition, considering contributions from (a) all phonon modes, and (c) only $a_1$-type phonon modes. Vibrational overlap functions (VOFs) for the ${}^3\!E \to {}^1\!A_1$ transition, considering contributions from (b) all phonon modes, and (d) only $a_1$-type phonon modes. Blue lines represent PL spectral functions and VOFs computed using the original partial Huang-Rhys factors, while red lines correspond to those computed with scaled partial Huang-Rhys factors. The experimental PL lineshape, taken from Ref.~\cite{alkauskas2014first}, is divided by $(E_{\mathrm{ZPL}} - \Delta)^3$ and normalized to one to derive the PL spectral function; $E_{\mathrm{ZPL}}$ is the energy of the zero-phonon line (ZPL).}
    \label{s-fig:spectral_fxns}
\end{figure*}

The computed VOFs are shown in Fig.~\ref{s-fig:spectral_fxns}, including both the PL lineshapes of the ${}^3\!E \to {}^3\!A_2$ transition and the VOFs of the ${}^3\!E \to {}^1\!A_1$ ISC transition. The results include VOFs computed considering all phonon modes as well as those restricted to the $a_1$-type phonon modes. For the PL of the ${}^3\!E \to {}^3\!A_2$ transition, a scaling factor of 1.57 applied to the computed HR factors of $a_1$-type phonon modes leads to a good agreement with experimental data~\cite{alkauskas2014first}. This scaling factor is derived from the ratio of the experimental total HR factor for $a_1$-type phonon modes--approximated as the experimental total HR factor (3.49~\cite{kehayias2013infrared}) minus the computed total HR factor for $e$-type phonon modes (0.59)--to the computed total HR factor for $a_1$-type phonon modes (1.85). The scaling factor accounts for the errors in the geometry relaxation of the ${}^3\!E$ state due to the use of the PBE functional in TDDFT calculations~\cite{razinkovas2021vibrational,jin2022vibrationally}. The same scaling factor is applied to the $a_1$-type phonon mode HR factors in the VOF calculation for the ${}^3\!E \to {}^1\!A_1$ ISC transition.

\section{Analysis of the Herzberg-Teller Effect}

The SOC matrix element $\lambda_\perp^{\mathrm{QDET}}$ is calculated at the equilibrium atomic geometry of the ${}^3\!A_2$ state and is assumed to remain constant for the calculation of ISC rates. To investigate the dependence of $\lambda_\perp^{\mathrm{QDET}}$ on atomic geometries, we compute $\lambda_\perp^{\mathrm{QDET}}$ using a 215-atom supercell at the equilibrium atomic geometries of the ${}^3\!A_2$ and ${}^1\!A_1$ states, with an active space of 40 orbitals. The computed values are 56.62 GHz and 54.48 GHz, respectively. Based on these results, a correction factor of $54.48 / 56.62 = 0.962$ is applied to $\lambda_{\perp,0}^{\mathrm{QDET}}$ in the ISC rate calculations.

\begin{figure*}
    \centering
    \includegraphics[width=16cm]{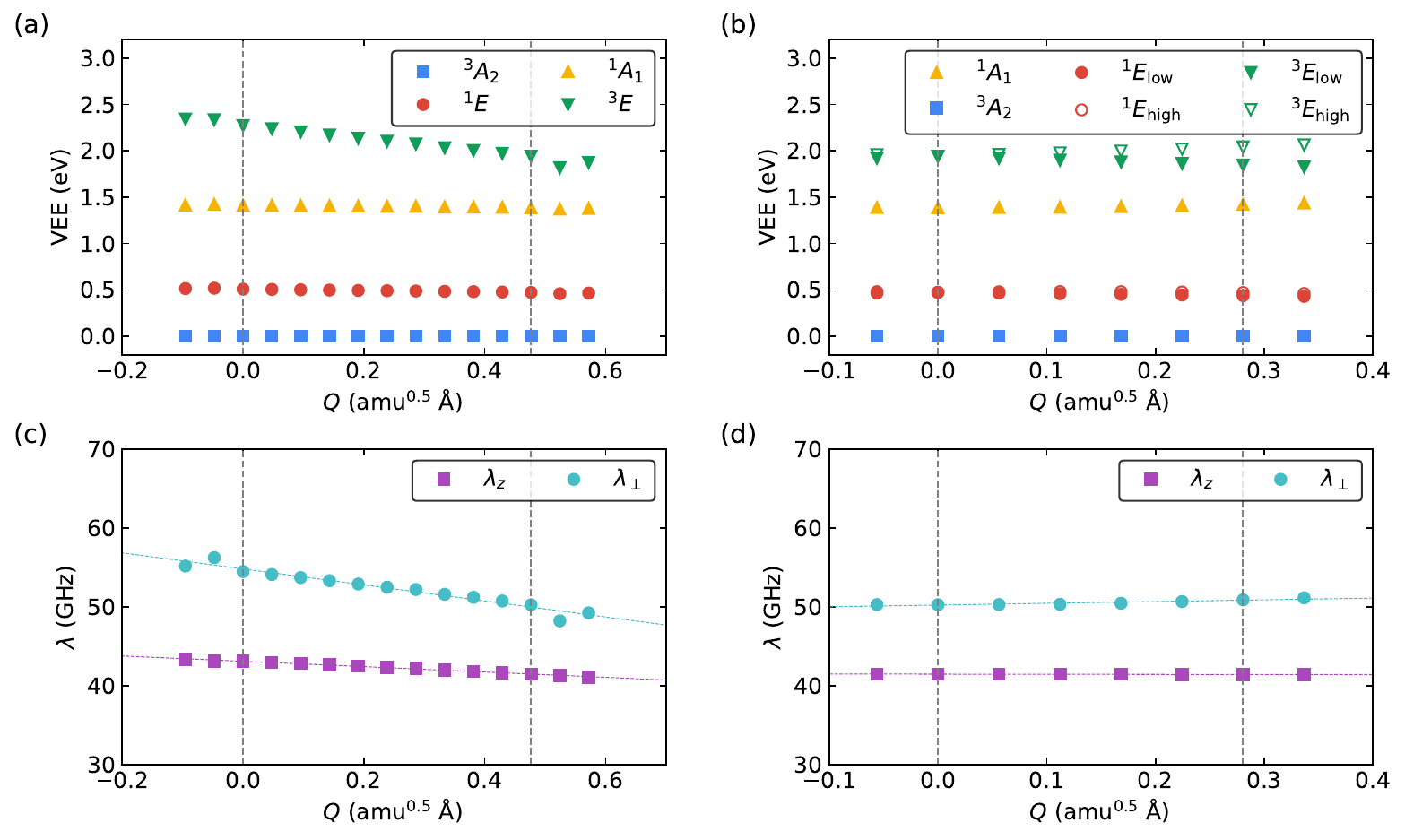}
    \caption{Vertical excitation energies (VEEs) of the ${}^1\!E$, ${}^1\!A_1$, and ${}^3\!E$ states relative to the ${}^3\!A_2$ ground state, along with SOC matrix elements ($\lambda_z$ and $\lambda_\perp$), are computed using the quantum defect embedding theory (QDET) with a 215-atom supercell and shown as functions of one-dimensional configuration coordinates ($Q$). In (a) and (c), the configuration coordinate corresponds to the linear path between the equilibrium atomic geometry of the ${}^1\!A_1$ state ($Q = 0.0$ amu$^{0.5}$ \AA) and the symmetrized equilibrium atomic geometry of the ${}^3\!E$ state ($Q = 0.477$ amu$^{0.5}$ \AA), reflecting the contribution of $a_1$-type phonon modes. In (b) and (d), the configuration coordinate corresponds to the linear path between the symmetrized equilibrium atomic geometry ($Q = 0.0$ amu$^{0.5}$ \AA) and the symmetry-broken equilibrium atomic geometry of the ${}^3\!E$ state ($Q = 0.281$ amu$^{0.5}$ \AA), capturing the contribution of $e$-type phonon modes. Symmetry breaking leads to the splitting of the degenerate ${}^1\!E$ state into two branches, denoted as ${}^1\!E_{\textrm{low}}$ and ${}^1\!E_{\textrm{high}}$. A similar splitting occurs for the ${}^3\!E$ state.}
    \label{s-fig:qdet_soc_derivative}
\end{figure*}

\begin{figure*}
    \centering
    \includegraphics[width=8cm]{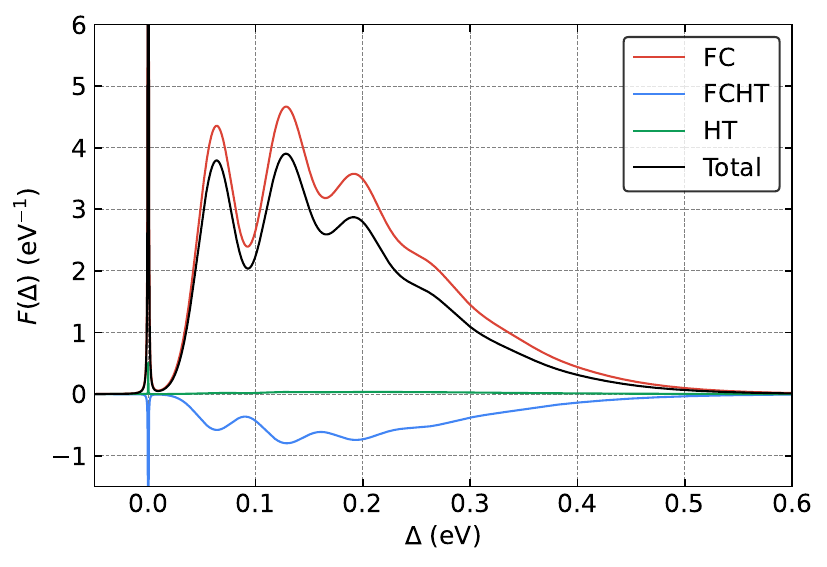}
    \caption{Vibrational overlap function (VOF) for the ${}^3\!E \to {}^1\!A_1$ transition, computed using the one-dimensional effective phonon approximation, as a function of the energy gap ($\Delta$) between ${}^3\!E$ and ${}^1\!A_1$. Individual contributions from the Franck-Condon (FC, red line), Franck-Condon Herzberg-Teller (FCHT, blue line), and Herzberg-Teller (HT, green line) terms are shown individually. The total VOF, including all contributions, is represented by the black line.}
    \label{s-fig:1d_spectral_fxn_HT}
\end{figure*}

Next, we investigate the dependence of $\lambda_\perp^{\mathrm{QDET}}$ on changes in atomic geometries from the ${}^1\!A_1$ to ${}^3\!E$ states. Two paths are constructed: the first path connects the equilibrium geometry of the ${}^1\!A_1$ state to the symmetrized equilibrium geometry of the ${}^3\!E$ state, which includes contributions only from the $a_1$-type phonon modes. The second path connects the symmetrized equilibrium geometry of the ${}^3\!E$ to its symmetry-broken equilibrium geometry, including contributions only from the $e$-type phonon modes. As shown in Fig.~\ref{s-fig:qdet_soc_derivative}, $\lambda_\perp^{\mathrm{QDET}}$ decreases linearly by approximately 8\% from the ${}^1\!A_1$ geometry to the symmetrized ${}^3\!E$ geometry along path 1, and by an additional 1\% from the symmetrized to the symmetry-broken ${}^3\!E$ geometry along path 2. Here, we consider the impact of the $a_1$-type phonon modes using a one-dimensional (1D) effective phonon approach~\cite{jin2021photoluminescence}, expressing $\lambda_\perp$ as a linear function of the mass-weighted displacement $Q$:
\begin{equation}
    \lambda_\perp (Q) = \lambda_\perp (Q = 0) + \dfrac{d \lambda_\perp (Q)}{d Q} \bigg|_{Q=0} Q,
\end{equation}
where $Q = 0$ corresponds to the ${}^1\!A_1$ geometry. The ISC rate at $T = 0$ K is then decomposed into three terms: the Franck-Condon (FC), Franck-Condon Herzberg-Teller (FCHT), and Herzberg-Teller (HT) terms~\cite{herzberg1933schwingungsstruktur,lin1974study}
\begin{equation}
    \Gamma = \Gamma_{\mathrm{FC}} + \Gamma_{\mathrm{FCHT}} + \Gamma_{\mathrm{HT}},
\end{equation}
or equivalently:
\begin{equation}
    \Gamma = \frac{4\pi}{\hbar} \left| \lambda_\perp (Q = 0) \right|^2 F_{\mathrm{Total}} (\Delta) = \frac{4\pi}{\hbar} \left| \lambda_\perp (Q = 0) \right|^2 \left[ F_{\mathrm{FC}} (\Delta) + F_{\mathrm{FCHT}} (\Delta) + F_{\mathrm{HT}} (\Delta) \right].
\end{equation}
These terms are defined as:
\begin{equation}
\begin{aligned}
    F_{\mathrm{Total}} (\Delta) &= F_{\mathrm{FC}} (\Delta) + F_{\mathrm{FCHT}} (\Delta) + F_{\mathrm{HT}} (\Delta),
\end{aligned}
\end{equation}
where
\begin{equation}
    F_{\mathrm{FC}} (\Delta) = \sum_j \left| \left \langle \chi_0 (Q) | \chi_j (Q + \Delta Q) \right\rangle \right|^2 \delta \left( \Delta - n_j \hbar\omega \right),
\end{equation}
\begin{equation}
    F_{\mathrm{FCHT}} (\Delta) = \frac{2 \frac{d \lambda_\perp (Q)}{d Q} }{\lambda_\perp (Q=0)} \sum_j \left \langle \chi_0 (Q) | \chi_j (Q + \Delta Q) \right\rangle \left \langle \chi_j (Q + \Delta Q) | Q | \chi_0 (Q) \right\rangle \delta \left( \Delta - n_j \hbar\omega \right), \\
\end{equation}
\begin{equation}
    F_{\mathrm{HT}} (\Delta) = \left|\frac{\frac{d \lambda_\perp (Q)}{d Q} }{\lambda_\perp (Q=0)}\right|^2 \sum_j \left| \left \langle \chi_0 (Q) | Q |\chi_j (Q + \Delta Q) \right\rangle \right|^2 \delta \left( \Delta - n_j \hbar\omega \right). \\
\end{equation}
Using the data from Fig.~\ref{s-fig:qdet_soc_derivative}(c), we set $\frac{\frac{d \lambda_\perp (Q)}{d Q} }{\lambda_\perp (Q=0)} = -0.162$ GHz / (amu$^{0.5}$ \AA), $\Delta Q = 0.597$ amu$^{0.5}$ \AA\ (calculated from the mass-weighted displacement between the ${}^1\!A_1$ and ${}^3\!E$ atomic geometries, $\Delta Q = 0.477$ amu$^{0.5}$ \AA, scaled by the square root of the 1.57 scaling factor derived in Sec.~\ref{s-sec:spectral_fxn}), and $\hbar\omega = 63$ meV~\cite{jin2022vibrationally} in the calculation of $F_{\mathrm{FC}}$, $F_{\mathrm{FCHT}}$, and $F_{\mathrm{HT}}$ by computing the overlap integrals recursively~\cite{ruhoff1994recursion}. The resulting VOFs are shown in Fig.~\ref{s-fig:1d_spectral_fxn_HT}.

We find that the HT term contributes negligibly to the total VOF, while the FCHT term contributes negatively, reducing the VOF intensity by approximately 20\%. Therefore, both the FCHT and HT terms, computed using this approach, are included in the VOF calculations discussed in the main text. It should be noted that the treatment of the FCHT and HT terms using the 1D effective phonon model is an approximation, valid for cases with large $\lambda_\perp (Q=0)$ and small derivative $\frac{d \lambda_\perp(Q)}{dQ}$. For scenarios involving large derivatives or cases where $\lambda_\perp (Q=0) = 0$, a more accurate approach would require explicit treatment of individual phonon modes~\cite{libbi2022phonon}. Our computational workflow can be readily adapted for such cases.

\section{Temperature Dependence of Intersystem Crossing Rates}
\begin{figure*}
    \centering
    \includegraphics[width=16cm]{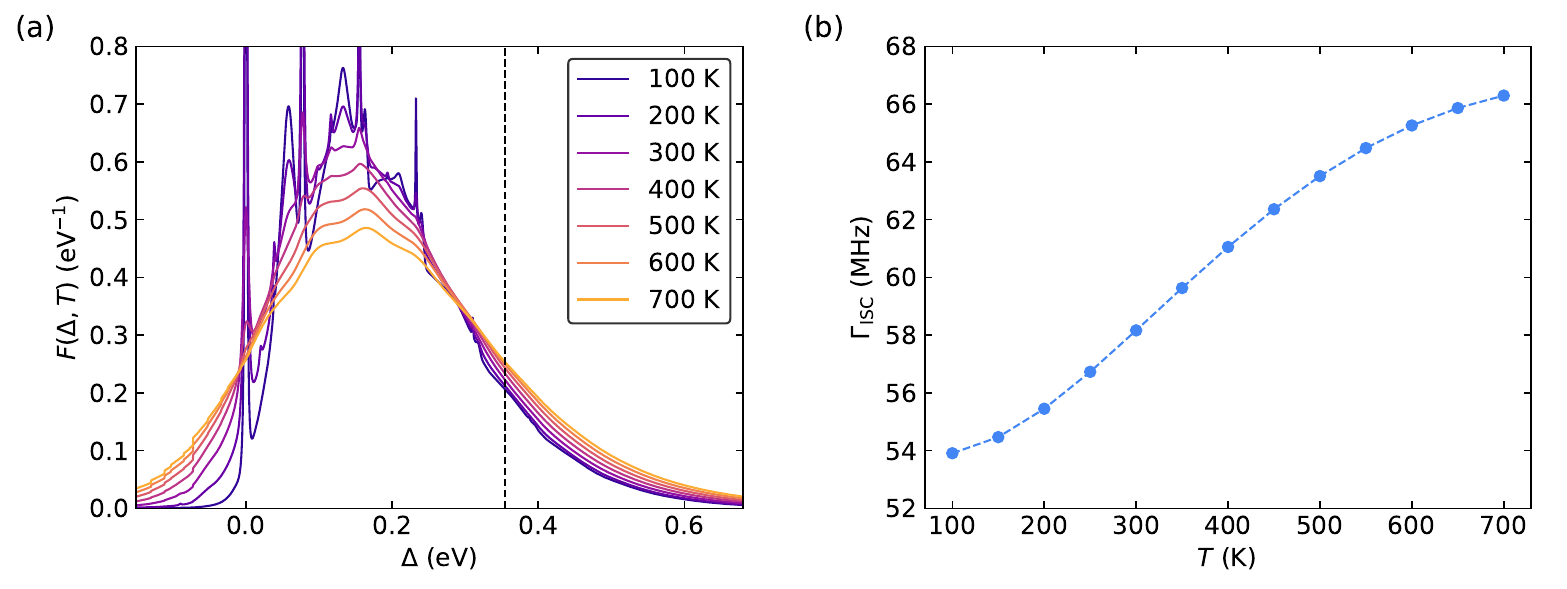}
    \caption{(a) Computed Vibrational overlap functions (VOFs) as a function of temperature, including the dynamic Jahn-Teller (DJT) and Herzberg-Teller (HT) effects. (b) Finite-temperature intersystem crossing (ISC) rates calculated using Eq.~\ref{s-eq:isc_temp}, with $\Delta = 0.355$ eV, which is the midpoint of the interval $[\Delta_-, \Delta_+]$, determined to match ISC rates at near-zero temperature (see main text).}
    \label{s-fig:temp_spectral_fxn}
\end{figure*}

The ${}^3\!E (|m_s| = 1) \to {}^1\!A_1$ ISC rate at finite temperatures can be computed as:
\begin{equation}
    \Gamma_{\mathrm{ISC}} (T) = \frac{4\pi}{\hbar} \left| \lambda_\perp \right|^2 F(\Delta, T),
    \label{s-eq:isc_temp}
\end{equation}
where
\begin{equation}
    F(\Delta, T) = \sum_{i} P_{i}(T) F_i(\Delta, T)
\end{equation}
is the VOF at temperature $T$. $P_i(T) = \frac{\exp\left(-\frac{\widetilde{E}_i}{k_BT}\right)}{\sum_j \exp\left(-\frac{\widetilde{E}_j}{k_BT}\right)}$ is the thermal distribution function for the $i$th vibronic level with energy $\widetilde{E}_i$, which is obtained by solving the DJT effective Hamiltonian Eq.~\ref{s-eq:eff_h}. The summation over all vibronic levels effectively captures the phonon-induced orbital averaging effect~\cite{goldman2015state}, yielding the averaged ISC rate from the $|m_s| = 1$ sublevels of ${}^3\!E$. The corresponding VOF, $F_i(\Delta, T)$, is computed as:
\begin{equation}
    F_i(\Delta, T) = \sum_{j} \left|c_{i,j} \right|^2 F^{a_1} \left(\Delta - n_j \hbar \omega_e + \widetilde{E}_i, T \right),
\end{equation}
where $c_{i,j}$ is the expansion coefficient of the $i$th vibronic level in terms of the $A_1$ electronic sublevel and the vibrational wavefunction $\chi_j$, with $n_j$ being the phonon number. The VOF $F^{a_1}(\Delta, T)$ accounts for the $a_1$-type phonon modes at temperature $T$, including the contributions of the HT effect. The temperature-dependent VOFs $F(\Delta, T)$ are shown in Fig.~\ref{s-fig:temp_spectral_fxn}. In the calculation of the temperature-dependent ISC rates, the VOF is evaluated at $\Delta = 0.355$ eV. This value represents the midpoint of the range $[0.334, 0.375]$ eV, determined by the intersection of the computed ISC rates with experimental results, as discussed in the main text.

\section{Comparison with Previous Theoretical Studies}

Here, we summarize the comparison of our results with those of previous theoretical studies on the ISC rates of the NV$^-$ center in diamond, including the studies by Goldman et al.~(2015)~\cite{goldman2015state,goldman2017erratum}, Thiering et al.~(2017)~\cite{thiering2017ab}, and Li et al.~( 2024)~\cite{li2024excited}.

The study by Goldman et al.~\cite{goldman2015state,goldman2017erratum} is not based on first-principles calculations. It assumes $\lambda_\perp = (1.2 \pm 0.2) \lambda_z$, where $\lambda_z$ is derived from the measured fine structure splitting of the ${}^3\!E$ state, divided by the Ham reduction factor to account for the DJT effect. Using $p=0.304$~\cite{thiering2017ab}, $\lambda_\perp$ is estimated to be $21.1 \pm 3.6$ GHz, while $p = 0.43$~\cite{goldman2017erratum} yields $\lambda_\perp = 14.9 \pm 2.5$ GHz. Both values are smaller than $\lambda_{\perp,0}^{\mathrm{QDET}} = 29.4 \pm 2.8$ obtained in this work. Additionally, Goldman et al. use the measured PL spectral function for the ${}^3\!E \to {}^3\!A_2$ transition to approximate the VOF $F_{A_1}(\Delta)$, without accounting for the DJT and HT effects. As a result, their estimated $F_{A_1}(\Delta)$ is larger than the value computed in this work (see Fig.~3 in the main text). By matching their results with the experimentally derived $\Gamma_{A_1}$ of $100.5 \pm 3.8$ MHz, Goldman et al. estimate a range of $\Delta$ of [0.321, 0.414] eV. This range encompasses the $\Delta$ range obtained in this work, due to the cancellation effect of their smaller $\lambda_\perp$ and higher $F_{A_1}(\Delta)$.

The study by Thiering et al.~\cite{thiering2017ab} adopts $\lambda_\perp = 56.3$ GHz, derived from DFT calculations with the HSE functional and the projector-augmented wave (PAW) method. This value is 1.9 times larger than $\lambda_{\perp,0}^{\mathrm{QDET}}$ obtained in this work, primarily due to the insufficient treatment of electron correlation effects in the many-body wavefunctions, leading to a 3.8-fold overestimation of the ISC rate. Furthermore, Thiering et al. use the VOF $F^{a_1}(\Delta)$ with a total HR factor $S=2.61$, obtained from constrained-occupation DFT ($\Delta$SCF) calculations, which exceeds the value $S=2.23$ determined in this work. This discrepancy contributes an additional 1.5- to 2-fold overestimation of the ISC rate, assuming $\Delta \in [0.3, 0.4]$ eV. Together, these approximations result in a total overestimation of the ISC rate by 5.7--7.6 times. While Thiering et al. acknowledge that their computed ISC rates are an order of magnitude larger than experimental values, they do not explicitly report the computed rates.

The study of Li et al.~\cite{li2024excited} reports $\lambda_\perp = 6.75$ GHz, which is 0.23 times the value $\lambda_{\perp,0}^{\mathrm{QDET}}$ determined in this work. This discrepancy likely arises from the inaccurate description of many-body wavefunctions in the cluster model, resulting in a 0.053-fold underestimation of the ISC rate. Additionally, Li et al. use $\Delta = 0.41$ eV obtained from CASPT2 calculations and a VOF value of 1.34 eV$^{-1}$, which overestimates $F_{A_1}(\Delta) + 2F_{E_{1,2}}(\Delta)$ obtained in this work at $\Delta = 0.41$ eV by approximately a factor of 3. Moreover, their ISC rate expression omits a factor of 2 (related to the definition of $\lambda_\perp$) but includes an additional geometrical degeneracy $g=3$ factor, resulting in a 1.5-fold adjustment. The computed ISC rate reported in their study is 29.86 MHz, which is $0.15$ times the experimental value ($\Gamma_{A_1} + 2\Gamma_{E_{1,2}} = 204.9$ MHz).

\renewcommand{\arraystretch}{1.5}
\begin{table*}[ht]
\caption{Vertical excitation energies (VEEs) of the ${}^3\!E$ and ${}^1\!A_1$ states relative to the ${}^3\!A_2$ ground state computed using various theoretical approaches. $\Delta_{\mathrm{VEE}}$ represents the energy difference. The adiabatic energy gap, $\Delta$, between the ${}^3\!E$ and ${}^1\!A_1$ states is calculated as $\Delta = \Delta_{\mathrm{VEE}} - E_{\mathrm{FC}, {}^3\!E} + E_{\mathrm{FC}, {}^1\!A_1}$, where $E_{\mathrm{FC}, {}^3\!E} = 0.256$ eV and $E_{\mathrm{FC}, {}^1\!A_1} = 0.016$ eV are the Franck-Condon (FC) shifts of the ${}^3\!E$ and ${}^1\!A_1$ states relative to the equilibrium atomic geometry of the ${}^3\!A_2$ state. These shifts were computed using TDDFT with the DDH functional~\cite{jin2023excited}. Intersystem crossing (ISC) rates $\Gamma_{A_1}$ and $\Gamma_{E_{1,2}}$ are computed for the value of $\Delta$ obtained by each theoretical approach, using $\lambda_{\perp,0}^{\mathrm{QDET}}$ and $F_{A_1/E_{1,2}}(\Delta)$ computed in this work. All energies are reported in eV, and the ISC rates are reported in MHz.}
\label{s-tab:vees}
\centering
\setlength{\tabcolsep}{8pt}
\begin{tabular}{c|c|c|c|c|c|c}
\hline\hline
 & VEE $\left(|{}^3\!E\rangle\right)$ & VEE $\left(|{}^1\!A_1\rangle\right)$ & $\Delta_{\mathrm{VEE}}$ & $\Delta$ & $\Gamma_{A_1}$ & $\Gamma_{E_{1,2}}$ \\ \hline
 TDDFT (PBE)~\cite{jin2023excited}  &  2.089  & 1.336 & 0.753 & 0.513 & $22 \pm 4$ & $11 \pm 2$ \\ 
 TDDFT (DDH)~\cite{jin2023excited}  &  2.372  & 1.973 & 0.399 & 0.159 & $432 \pm 82$ & $111 \pm 21$ \\ 
QDET (DC2025, PBE)~\cite{chen2025qdet}  & 2.008 & 1.475 & 0.533 & 0.293 & $180 \pm 34$ & $92 \pm 17$ \\
QDET (DC2025, DDH)~\cite{chen2025qdet}  & 2.235 & 1.589 & 0.646 & 0.406 & $67 \pm 13$ & $32 \pm 6$ \\
 QDET (HFDC, RPA)~\cite{ma2021qdet} & 1.921 & 1.376 & 0.545 & 0.305 & $173 \pm 33$ & $77 \pm 15$ \\
 NEVPT2-DMET~\cite{haldar2023local} & 2.31 & 1.56 & 0.75 & 0.51 & $23 \pm 4$ & $11 \pm 2$ \\
 CI-cRPA~\cite{bockstedte2018ab} & 2.02 & 1.41 & 0.61 & 0.37 & $103 \pm 20$ & $44 \pm 8$ \\
 CASSCF (C$_{85}$H$_{76}$N$^-$)~\cite{bhandari2021multiconfigurational} & 2.14 & 1.60 & 0.54 & 0.30 & $172 \pm 33$ & $84 \pm 16$ \\
 CASSCF (C$_{33}$H$_{36}$N$^-$)~\cite{li2024excited} & 2.30 & 1.96 & 0.34 & 0.10 & $448 \pm 85$ & $51 \pm 10$ \\
 CASPT2 (C$_{33}$H$_{36}$N$^-$)~\cite{li2024excited} & 2.22 & 1.57 & 0.65 & 0.41 & $65 \pm 12$ & $30 \pm 6$ \\
 ppRPA (PBE)~\cite{li2024accurate} & 1.95 & 1.67 & 0.28 & 0.04 & $375 \pm 72$ & $0.19 \pm 0.04$ \\
 ppRPA (B3LYP)~\cite{li2024accurate} & 2.10 & 1.97 & 0.13 & $-0.11$ & & \\
 $GW$-BSE~\cite{yu2024gpu} & 2.379 & 1.169 & 1.21 & 0.97 & $0.06 \pm 0.01$ & $0.027 \pm 0.005$ \\
 Expt.~\cite{goldman2015phonon,goldman2015state,goldman2017erratum} & & & & $[0.321,0.414]$ & $100.5 \pm 3.8$ & $52.2 \pm 2.0$ \textsuperscript{a} \\
 \hline\hline
\end{tabular}
\vspace{1mm}

\begin{minipage}{\textwidth}
\raggedright
\hspace{0mm}\textsuperscript{a} The relative uncertainty of $\Gamma_{E_{1,2}}$ is assumed to the same as that of $\Gamma_{A_1}$.
\end{minipage}
\end{table*}

\begin{figure*}
    \centering
    \includegraphics[width=10cm]{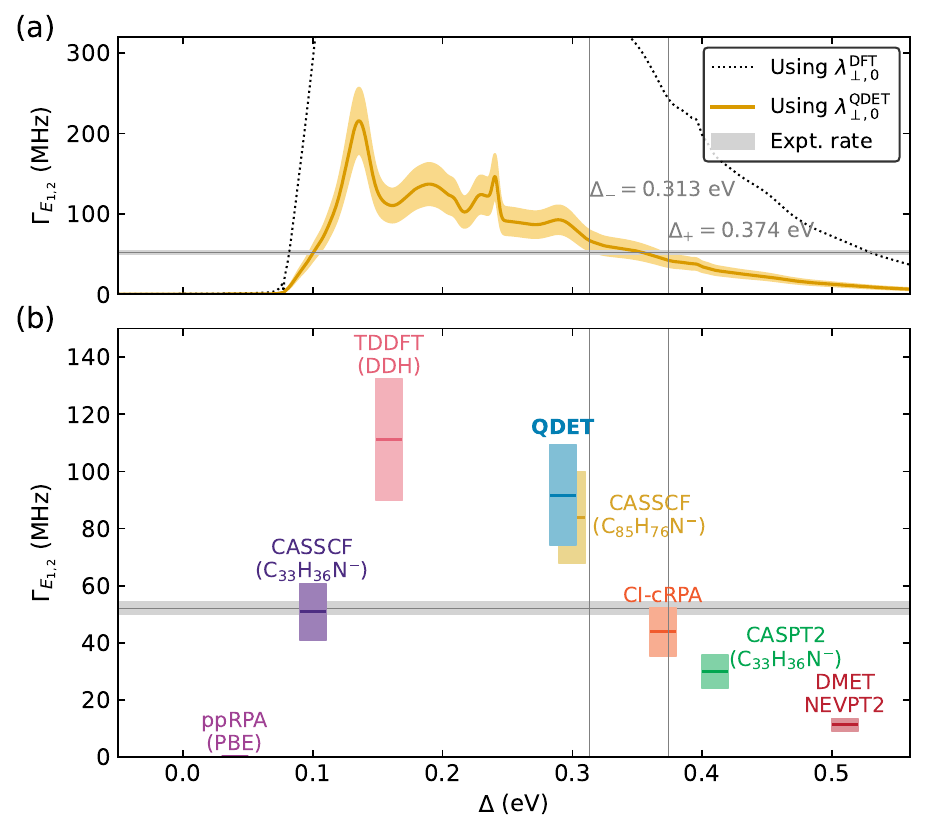}
    \caption{Intersystem crossing (ISC) rates $\Gamma_{E_{1,2}}$. (a) Computed $\Gamma_{E_{1,2}}$ as a function of the energy gap ($\Delta$) between the ${}^3\!E$ and ${}^1\!A_1$ states. The red shaded area represents the uncertainty in $\Gamma_{E_{1,2}}$ from the computed SOC matrix element $\lambda_{\perp,0}^{\text{QDET}}$. $\Delta_-$ and $\Delta_+$ correspond to the lower and upper bounds of $\Delta$, respectively, determined by the intersection of the computed $\Gamma_{E_{1,2}}$ with experimental results~\cite{goldman2015phonon}. An uncertainty of 2.0 GHz is added to the measured $\Gamma_{E_{1,2}}$ rate, based on the assumption that the relative uncertainty of $\Gamma_{E_{1,2}}$ is the same as that of $\Gamma_{A_{1}}$. $\Gamma_{E_{1,2}}$ computed using $\lambda_{\perp,0}^{\text{DFT}}$ is shown as a dotted line for comparison. (b) $\Gamma_{E_{1,2}}$ calculated using $\lambda_{\perp,0}^{\mathrm{QDET}}$ and the vibrational overlap function $F_{E_{1,2}}(\Delta)$ derived in this work, and $\Delta$ values from various theoretical methods~\cite{jin2023excited,chen2025qdet,ma2021qdet,haldar2023local,bockstedte2018ab,bhandari2021multiconfigurational,li2024excited,li2024accurate}. The value of $\Delta$ includes Franck-Condon shifts of the ${}^3\!E$ and ${}^1\!A_1$ states as obtained from TDDFT~\cite{jin2023excited}. A visualization width of 0.02 eV is applied to each $\Delta$. Detailed data are provided in Table~\ref{s-tab:vees}.}
    \label{s-fig:gamma_e12}
\end{figure*}

\section{Experimental Details}

Our custom-built confocal microscope set-up is designed to image and probe NV centers. We use a nonpolarized 520 nm laser (LABSelectronics DLnsec), which is internally modulated for pulsed measurements to achieve subnanosecond fall times. The laser is focused on the diamond sample with an Olympus MPLFLN100x objective, featuring a numerical aperture (NA) of 0.9 and 85\% transmission across the visible spectrum. NV photoluminescence is separated from the laser using a 550 nm longpass dichroic mirror and filtered through a 655 nm longpass before detection with an avalanche photodiode (PerkinElmer SPCM-AQR-16). The laser is scanned across the sample using a fast-steering mirror (Newport Corp.) to locate single NV centers. NV centers are confirmed via optically detected magnetic resonance (ODMR) by observing spin contrast. Spin rotations are performed using a vector signal generator (SRS SG396), with microwave signals amplified (Mini Circuits ZHL-16W-43+) and delivered to the sample via a 10-micron diameter gold wire. A permanent magnet (K\&J Magnetics) is mounted on a custom goniometer with a two-axis stage (Newport MFA-CC) for precise alignment. All equipment timing is synchronized using an arbitrary waveform generator (Swabian Instruments Pulse Streamer 8/2). Temperature control between 100K and room temperature is managed using a Montana Instruments Nanoscale Workstation.

To probe the spin-dependent optical lifetime of the NV, we performed Time-Correlated Single Photon Counting (TCSPC) using a time-tagger module (Swabian Instruments Time Tagger Ultra). The pulse sequence follows a procedural series of three pulses: (re)initialization, spin-state preparation, and spin-state readout. The series of pulses are for both $m_s = -1$ (RF on) and $m_s = 0$ (RF off) and is repeated to build histogram counts on a respective time-tagger counter-channel that is connected to our APD output. From this procedure, we extract the spin-dependent optical lifetime of the NV at varying target temperatures.

\section{Fitting Optical-Lifetimes}

In order to extract the optical lifetimes, we perform a bi-exponential fit on the falling edge of the photon counts ($f$) by time ($t$) data:

\begin{equation*}
f(t)=\frac{1}{R+1}e^{-t/\tau_1}+\frac{R}{R+1}e^{-t/\tau_2}+C 
\end{equation*}

Where R is the ratio between the $\text{NV}^{-}$and $\text{NV}^{0}$ charge state populations for our sample, $\tau_1$ and $\tau_2$ are the two lifetimes for each charge state, $C$ represents the background noise which is comprised of dark counts, thermal noise and electric noise.
We then use a Bayesian Monte-Carlo Markov chain (MCMC) sampling algorithm to fit the falling edge of the readout pulse to extract the two optical lifetimes of our NV center defect in the neutral (NV0) charge state or negative charge state (NV-) with spin states $m_s = 0,-1$. To circumvent the difficulties due to the correlations of the parameters ~\cite{bretthorst2005exponential} we use the NUTS (No U-Turn Sampler) sampler~\cite{hoffman2011nuts} to obtain our samples implemented in the package PyMC~\cite{abril2023pymc} for python. Conventional MCMC algorithms such as random-walk Metropolis ~\cite{metropolis1949montecarlo} and Gibbs sampling~\cite{geman1984gibbs} suffer from long convergence times when sampling from correlated parameters, making them unsuitable for fitting to our model, as the population ratios and lifetimes are correlated to some extent. Therefore, in the interest of cost efficiency we choose to use a Hamiltonian Monte-Carlo approach ~\cite{neal1996hmc}, as it circumvents such issues by informing each sampling step via the first-order gradient. An optimized version of conventional HMC is given by NUTS, which reduces the computational cost further by halting the algorithm at the first onset of step retracing. 
To further improve the speed of our sampling, we use the numpyro~\cite{phan2019numpyro} implementation of the NUTS sampler, which reduces sampling overhead by using JIT compilation~\cite{aycock2003jit} on the GPU. The sampling is performed using 8 Markov chains and 5000 samples for each temperature, parameters similar to that of~\cite{toyli2012measurement}. For the prior distributions of all parameters, we choose informative normal distributions. For the lifetime parameters with the means selected according to previous experimental studies, and the standard deviation being chosen according to the theoretical model’s predictions, 0.5 ns. For the ratio parameter the mean chosen was 1, representing the state of no knowledge on the charge state populations, with a standard deviation of 0.5. The background parameter’s prior distribution was assumed to have a mean of 0 with a standard deviation of 0.1 to minimize the distortion of the other parameter values. 

To obtain the optimal interval for fitting, we repeat our Bayesian inference procedure for multiple different truncated intervals. After selecting the interval from the first onset of decay to the first plateau, we repeat our bi-exponential fit for different truncations of the head and extensions of the tail of our interval. We record each parameter and their error values along with the $R^2$ of the fit. We then select the optimal lifetime by the following criteria. First, we reject any fits for which the R parameter is negative, as this indicates an inadequate truncation. We then pre-select all the fits for which $R^2 > 0.98$, and $\hat{R} < 1.05$~\cite{gelman1992rhat} then select the fit with the lowest error for $\tau_2$, which gives us our optimal lifetime value. This is streamlined by an automation script, which sweeps the selected range and saves the parameter values for post-selection. To detect the falling edges of the peaks, we utilize a peak finding algorithm implemented in the scipy python package~\cite{virtanen2020scipy} to extract the broad ranges and then adjust by applying the breakpoint detection algorithm implemented in the ruptures package for python~\cite{truong2018ruptures}.

\bibliography{SM}